\documentclass[onecolumn,secnumarabic,amssymb, nobibnotes, aps, pra,10pt]{revtex4-2}

\usepackage{graphicx}
\usepackage{dcolumn}
\usepackage{bm}
\usepackage{hyperref}
\usepackage{ulem}

\graphicspath{{./fig/}}
\usepackage{color}
\usepackage{xcolor}

\begin{document}

\preprint{APS/123-QED}

\title{Dynamics of a reactive spherical particle falling in a linearly stratified fluid}

\author{Ludovic Huguet}
\email{ludovic.huguet@irphe.univ-mrs.fr}
\author{Victor Barge-Zwick}%
\author{Michael Le Bars}
\affiliation{CNRS, Aix Marseille Univ, Centrale Marseille, IRPHE, Marseille, France}%

\date{\today}

\begin{abstract}
Motivated by numerous geophysical applications, we have carried out laboratory experiments of a reactive (i.e. melting) solid sphere freely falling by gravity in a stratified environment, in the regime of large Reynolds $(Re)$ and Froude numbers. We compare our results to non-reactive spheres in the same regime. First, we confirm for larger values of $Re$, the stratification drag enhancement previously observed for low and moderate $Re$ \cite[e.g.][]{mm2020}. We also show an even more significant drag enhancement due to melting, much larger than the stratification-induced one. We argue that the mechanism for both enhancements is similar, due to the specific structure of the vorticity field sets by buoyancy effects and associated baroclinic torques, as deciphered for stratification by Zhang et al.~\cite{zmm2019}. Using particle image velocimetry, we then characterize the long-term evolution (at time $t \gg 1/N$ with $N$ the Br\"unt-V\"ais\"al\"a frequency) of the internal wave field generated by the wake of the spheres. Measured wave field is similar for both reactive and inert spheres: indeed, each sphere fall might be considered as a quasi impulsive source of energy in time and the horizontal direction, as the falling time (resp. the sphere radius) is much smaller than $N$ (resp. than the tank width). Internal gravity waves are generated by wake turbulence over a broad spectrum, with the least damped component being at the Br\"unt-V\"ais\"al\"a frequency and the largest admissible horizontal wavelength. About 1\% of the initial potential energy of each sphere is converted in to kinetic energy in the internal waves, with no significant dependence on the Froude number over the explored range.

\end{abstract}

\maketitle

\section{Introduction}

Particles settling or rising in a stratified fluid have been widely studied in the previous decades because of numerous applications ranging from industrial processes to geophysics (see the broad review of Magnaudet et al.~\cite{mm2020}). Examples include marine snow, plankton, and Lagrangian floats \cite{d2003} in the ocean, as well as dust and aerosols in the atmosphere. 
Moreover, moving reactive particles, meaning particles exchanging heat and mass with the surrounding medium, are encountered in many geophysical contexts such as ice crystallizing in the atmosphere \cite{ckudkl2019}, water droplet condensing or evaporating in clouds, iron or oxide crystals solidifying in planetary cores \cite{had2006,bsn2016,zndl2019}. The sedimentation of these reactive particles often occurs in a stably stratified layer, as for instance at the top or bottom of liquid planetary cores due to a combination of thermal, pressure and chemical gradients \cite{mdra2019}. The fall of reactive particles may strongly interact with the stratified surrounding environment, and the associated dynamics are the focus of our present study, using a generic, laboratory, experimental model. These dynamics include both the fall of the particle and its wake, but also the internal wave field that it generates and that then persists for a long period of time.

\begin{figure*}
\includegraphics[]{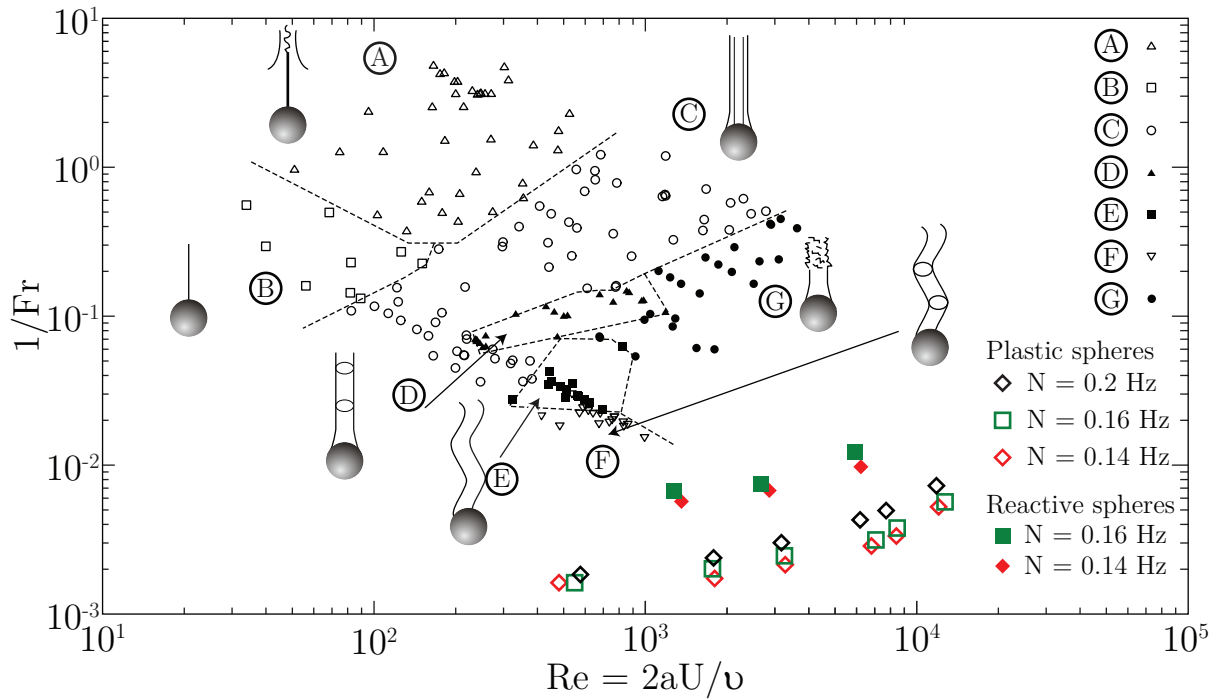}
\caption{Diagram of the 7 regimes of wakes observed in the experiments or numerical studies \cite{mm2020}. The solid and empty large markers denote our experiments with the reactive and plastic spheres respectively. Modified from Fig.~4a of Hanazaki et al.~\cite{hko2009a} with the permission of Cambridge University Press.}
\label{fig:regimediagram}
\end{figure*}

Experimental studies have reported a drag increase for inert particles falling through a sharp density gradient \citep{smf1999,aaam2004,cflmrp2009,cflmm2010,pm2018,vvl2019}. In a linearly stratified fluid, Yick et al.~\cite{ytps2009} showed that the drag coefficient can be enhanced by a factor 3 compared to the one in a homogeneous fluid for small Reynolds number ($Re = \frac{2aU}{\nu}$ with $U$ the falling velocity, $a$ the sphere radius and $\nu$ the fluid kinematic viscosity).
A lot of numerical simulations of a sphere falling in a stratified layer have been done for small and moderate Reynolds numbers \cite{thocv2000,ytps2009,mcm2018,zmm2019,lfl2019}. Direct numerical simulations and asymptotic approaches have shown the enhancement of the drag due to the stratification. Doostmohammadi et al.~\cite{dda2014} also examined the drag enhancement in transient settling. 
Zhang et al. \cite{zmm2019} challenged the canonical view of the enhancing drag coefficient in a stratified layer, commonly attributed to the additional buoyancy force resulting from the dragging of light fluid by the falling body. They rigorously examined the physical mechanism implied in the drag increase and showed that for high Prandtl number the added drag is mostly due a modification of the vorticity field induced by the baroclinic torque on the surface of the sphere. Others experimental and numerical studies focused on the flow regime past the sphere settling \cite{thocv2000,hko2009b,hko2009a,hny2015}. Hanazaki et al.~\cite{hko2009a} investigated the characteristics (length and radius) of the associated jet and its behavior for a large range of Reynolds and Froude numbers ($Fr = \frac{U}{Na}$ with $N$ the Br\"unt-V\"ais\"al\"a frequency). 
They found seven regimes corresponding to various jet structures (see Fig.~\ref{fig:regimediagram}). Hanazaki et al.~\cite{hko2009b} examined the effect of the molecular diffusion and showed that the jet widens when diffusivity increases and that the stratification is then less effective.

The falling or rising paths of spherical or non-spherical bodies have been largely studied (see the review of Ern et al. \cite{erfm2012}). The trajectory of a falling sphere depends on the Reynolds number and the density ratio ($m^\star=\rho_{sphere}/\rho_{fluid}$) and can be oblique, periodic or chaotic \cite{hw2010,erfm2012}. For instance, at $m^\star \gg 1$ and high Reynolds number, falling paths are chaotic \cite{jdb2004}, but become periodic at $m^\star \sim 1$ \cite{mj1964,hw2010}. However, experimental studies of the falling or rising path are difficult as the noise background in the fluid can dramatically change the instabilities occurring on the path \cite{jdb2004,hw2010}. To the best of our knowledge, there is no exhaustive study of the path of a freely falling sphere in a stratified layer.

In a stratified layer, the vertical motion of a buoyant fluid or particle may generate internal waves \cite{mpfm1973,c1978}, as for instance in the Earth's atmosphere \cite{as2010} and in astrophysical environments \cite{zcs2018}. Mowbray and Rarity \cite{mr1967} have been the firsts to describe internal waves associated with the settling of a sphere through a stratified layer, in the regime of small Reynolds and Froude numbers. The velocity field during the fall of the sphere is explained by the linear theory of internal waves \cite{thocv2000,hko2009a,oah2017}, which governs the radius and the length of the jet behind the sphere. Most of the studies regarding the generation of internal waves by a turbulent wake are focused on a horizontally towed sphere in a linearly stratified layer. Two regimes of waves are described: Lee waves and random internal waves \citep{bch1993,lbf1993,rdz2020}, the second regime being dominant for $Fr> 4$ \citep{cbbhp1991,bch1993}. To the best of our knowledge, no previous study has examined the internal waves produced by the fall of a sphere in the regime of large Reynolds and Froude numbers.

In this paper, we investigate the dynamics of a sphere falling in a linearly stably stratified fluid while melting, focusing on the drag coefficient and the associated internal waves generation. To decipher the effects of melting from the ones due to stratification, we also investigate the dynamics of inert (plastic) spheres. Our experiments have a range of Froude and Reynolds numbers of $80<Fr<600$ and $400<Re<15000$ respectively, i.e. relatively unexplored in regards to the previous experiments as shown in Fig.~\ref{fig:regimediagram} (see Magnaudet et al. \cite{mm2020} for a review). In Section~\ref{sec:method}, we describe the experimental setup, and we define the physical parameters. In Section~\ref{sec:melting}, we present a model for understanding the melting rate of our spheres, and we compare it with experimental results. We describe the falling behavior of the spheres, and we calculate the drag coefficient for melting and plastic spheres in Section~\ref{sec:fallingdrag}. Section \ref{sec:waves} presents our results on the internal waves generated by both types of spheres settling into a stratified layer. In Section \ref{sec:discussion} we discuss our main results and their implications.

\section{Method}
\label{sec:method}

\begin{figure}
\includegraphics[]{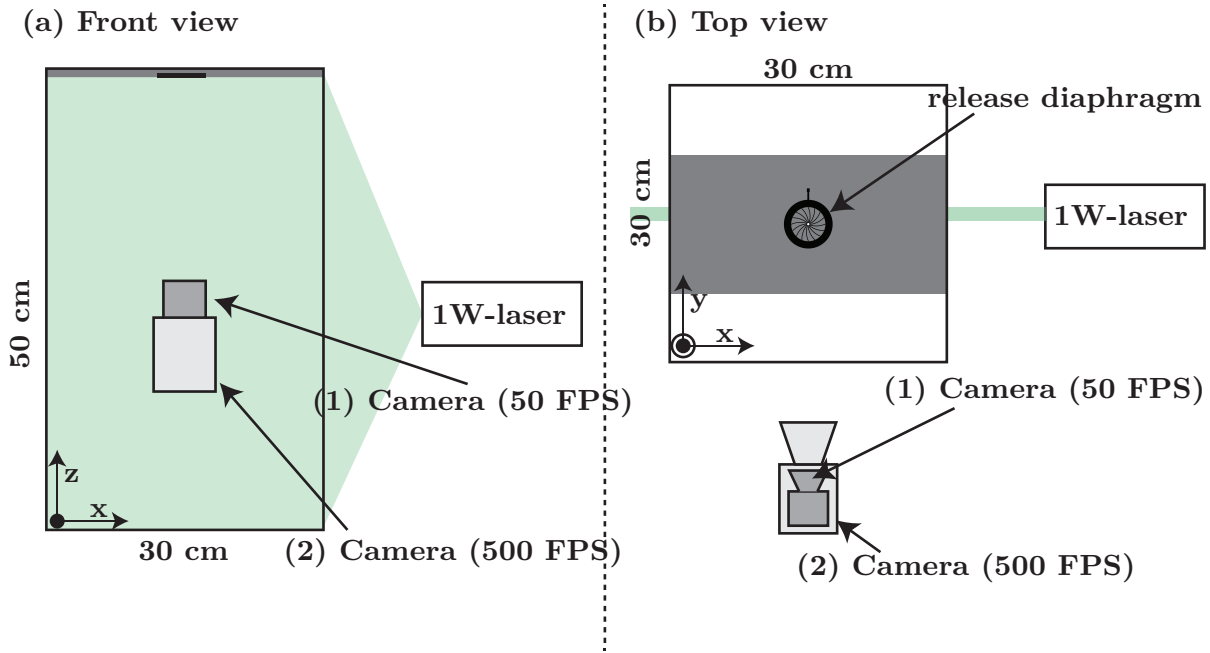}
\caption{Schema of the experimental set-up for the small tank. The 50~FPS-camera (1) allows us to perform PIV measurements over the whole tank. We use a 50 mm-lens on the 500~FPS-camera to track the rapid dynamics of the sphere wake in a reduced domain. A green laser of 1W is used here.}
\label{fig:schema}
\end{figure}
\begin{figure}
\includegraphics[]{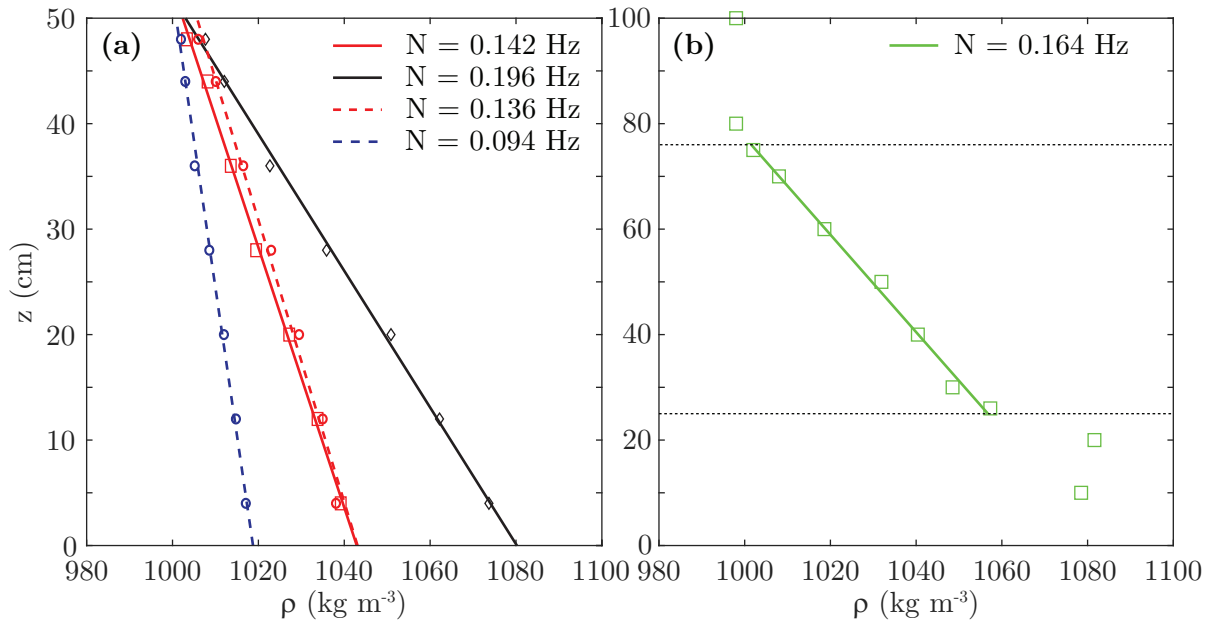}
\caption{(a) Density profile of the four different linear stratifications in the small tank. One stratification ($N = 0.142$~Hz) has been used with backlighting to track the sphere position (red solid line and squares). Two different stratifications ($N = 0.094 \text{ and } 0.136$~Hz) have been used for PIV measurements (dashed lines and circles). The last stratification ($N = 0.196$~Hz) has been used with the two measurement techniques (black line and diamonds). (b) Density profile in the large tank, only used with back-lighting technique: the linear stratification, of the same depth as in the small tank, is enclosed between the top and bottom homogeneous layers, with a noticeable density jump around the bottom interface.}
\label{fig:strati}
\end{figure}

The experiment is carried out in a $30\times 30\times 50$~cm tank filled with linearly stably stratified, salty water using the double-bucket technique \cite{o1965,eh2009} (Fig.~\ref{fig:schema}). A larger tank ($45\times 45\times 100$~cm) is used in one set of experiments to investigate possible wall effects. We impose a linear stratification which is characterized by the Br\"unt-V\"ais\"al\"a frequency $N$ in Hz 
\begin{equation}
N=\frac{1}{2\pi}\sqrt{-\frac{g}{\rho}\frac{\partial \rho}{\partial z}}
\label{brunt}
\end{equation}
where $g$ is the gravitational acceleration, $\rho$ is the mean density of the fluid, and $\frac{\partial \rho}{\partial z}$ is the density gradient. In our tanks, we set a Br\"unt-V\"ais\"al\"a frequency between 0.1 and 0.2~Hz. Fig. \ref{fig:strati} shows the five different stratifications used in our experiments. We have measured the density every 8~cm by microsampling and then using a portable density meter (Anton-Paar DMA 35). To release the sphere without initial shear or vertical velocity, we use an iris diaphragm with 16 leaves mounted on a support and disposed on the water surface, before being carefully opened (Fig.~\ref{fig:schema}). 
We perform particle image velocimetry (PIV) with silver-coated particles (diameter of 10~$\mu$m) to track the fluid motion in the whole tank at a frame rate of 50~FPS, as well as a zoom of the turbulent wake over a window of $12.3 \times 15~$~cm$^2$ at 500~FPS. In both cases, we typically use $32\times 32$~pixels boxes with $50 \%$ overlap. Independently of the PIV measurements, we also perform experiments with a back-lighting to track the real-time two-dimensional position of the spheres. The post-processing of these images produces better results for the sphere velocity, so for the drag coefficient. Indeed, the high-Reynolds number of our experiments implies a 3D motion of the falling spheres, which do not stay in the laser sheet plane. 

\begin{figure}
\includegraphics[]{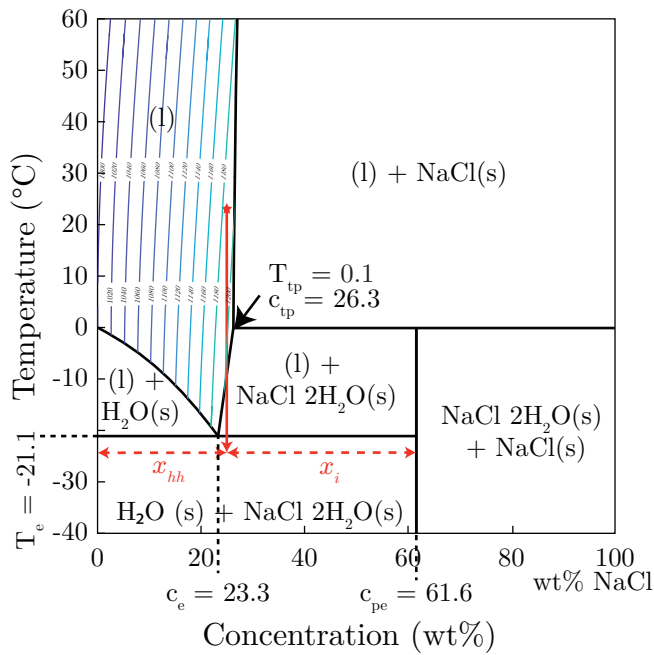}
\caption{Phase diagram of binary mixtures of water (H$_2$O) and sodium chloride (NaCl). (s) and (l) denote the solid and the liquid phases, respectively. The subscripts $_e$, $_{pe}$ and $_{tp}$ denote the eutectic, peritectic, and triple point of the phase diagram, respectively. The NaCl 2H$_2$O phase is a hydrated salt and is called hydrohalite. The fluid with a salt concentration of 25~wt\% (red star) is cooled below the eutectic temperature (vertical red arrow) where a solid mixture is formed by pure ice and hydrohalite (horizontal dotted red arrows). Blue color lines indicate the liquid density.}
\label{fig:phasediagram}
\end{figure}

Our reactive spheres are molded in spherical molds 3D-printed with three different radii (14, 7.9, 5~mm). By image analysis, we measure the ratio between the minor axis and the major axis of the falling spheres, which denotes their sphericity. For reactive spheres, this ratio is about $0.9\pm0.05$. The difficulty of making a proper sphere is due to (i) the small hole required to pour the liquid solution and (ii) the melting during the un-molding. The major axis is mostly perpendicular to gravity. To mold the reactive spheres, a liquid solution with $c_{0} = $~25~wt$\%$~NaCl is trapped in the spherical mold and cooled from ambient temperature to about $-23 ^\circ$C, below the eutectic temperature (see the red star arrow in Fig.~\ref{fig:phasediagram}). The mass composition of each sphere is then (see the horizontal red arrows in Fig.~\ref{fig:phasediagram}) 
\begin{equation}
x_i = \frac{c_{pe}-c_{0}}{c_{pe}} = 59.42\%~\text{of ice}
\label{eq:xi}
\end{equation}
and
\begin{equation}
x_{hh} = 1-\frac{c_{pe}-c_{0}}{c_{pe}} = 40.58\%~\text{of hydrohalite,}
\label{eq:xhh}
\end{equation}
where $c_{pe}$ is the concentration of the peritectic point. The density of the ice-hydrohalite mixture sphere $\rho_m$ is theoretically defined by 
\begin{equation}
\rho_{m,th}=\rho_{ice} (1-\Phi_{hh})+\rho_{hh} \Phi_{hh} = 1114~\text{kg\,m}^{-3},
\label{eqdensity}
\end{equation}
where $\rho_{ice}$ and $\rho_{hh}$ are the density of ice and hydrohalite, respectively (Table~\ref{tab_values}). $\Phi_{hh}=x_{hh}/(x_{hh}+(1-x_{hh})\frac{\rho_{hh}}{\rho_{ice}})$ is the volume fraction of hydrohalite. This value is slightly below our experimental measurements giving $\rho_{m,exp} = 1142 \pm 15~\text{kg\,m}^{-3}$, based on two different methods: first, we measured the mass of several spheres and the volume change induced by immersing them in water; second, we measured the equilibrium position of some spheres in a stratified fluid with a density linearly changing from 1096 to 1160~$\text{kg\,m}^{-3}$. The disagreement might essentially come from non-ideal conditions while making the sphere (e.g. non instantaneous cooling of the fluid below the eutectic). In the following sections, we thus use the $\rho_{m,exp}$ value.

In addition to those home-made reactive spheres, we also use spheres of \textit{Torlon} (polyamide-imide) or PVC (Polyvinyl chloride) with radius 1.6, 3.2, 4.8, 7.9, 9.5, 14.3~mm and density between 1300 and 1430 kg\,m$^{-3}$ to isolate the effect of melting on the falling sphere dynamics.

\section{Melting of our reactive spheres}
\label{sec:melting}
\subsection{Theoretical model}
\label{subsec:meltmodel}
\begin{table*}
\caption[]{\label{tab_values} Physical property values used in the theoretical calculations.}
\begin{ruledtabular}
\begin{tabular}{l|ccccc} 
Quantity & Symbol & \multicolumn{2}{c}{Value} & Unit & Reference \\
\colrule
Melting temperature of ice & $T_{ice}$ & \multicolumn{2}{c}{0}& $^\circ$C & \cite{s1983b} \\
Eutectic temperature & $T_{e}$ & \multicolumn{2}{c}{-21.1}& $^\circ$C &\cite{s1983b} \\ 
Triple point temperature & $T_{tp}$ & \multicolumn{2}{c}{0.1}& $^\circ$C &\cite{s1983b} \\ 
Eutectic composition & $c_{e}$ & \multicolumn{2}{c}{23.3} & wt\% &\cite{s1983b} \\ 
Peritectic composition & $c_{pe}$ & \multicolumn{2}{c}{61.6} & wt\% &\cite{s1983b} \\ 
Triple point composition & $c_{tp}$ & \multicolumn{2}{c}{26.33} & wt\% & \cite{s1983b}\\ 
Density of ice at -23$^\circ$C & $\rho_{ice}$ & \multicolumn{2}{c}{920}& kg\,m$^{-3}$ & \cite{m2007a} \\
Density of hydrohalite at -10$^\circ$C & $\rho_{hh}$ & \multicolumn{2}{c}{1610}& kg\,m$^{-3}$ & \cite{l1959,kyh2003} \\
Density of pure water at 25$^\circ$C & $\rho_{w}$ & \multicolumn{2}{c}{997}& kg\,m$^{-3}$ & \cite{k1975b,jh1992,mk2007} \\ 
Density of salty water at 0$^\circ$C and $c=25$~wt\%& $\rho_{0}$ & \multicolumn{2}{c}{1198.5}& kg\,m$^{-3}$ & see Appendix~\ref{sec:appendix:melting} \\
Latent heat of crystallization of ice & $L_{ice}$ & \multicolumn{2}{c}{334}& kJ\,kg$^{-1}$ & \cite{y1981} \\
Enthalpy of dissolution of NaCl & $L_{NaCl}$ & \multicolumn{2}{c}{66.39}& kJ\,kg$^{-1}$ & \cite{p1965} \\
Enthalpy of dissociation of hydrohalite & $L_{hh}$ & \multicolumn{2}{c}{7.73}& kJ\,kg$^{-1}$ & \cite{doy2017} \\
Heat capacity of water & $C_{p}^{l}$ & \multicolumn{2}{c}{4184}& J\,K$^{-1}$\,kg$^{-1}$ & \\
Thermal conductivity of the hydrohalite-ice sphere & $k_m$ & \multicolumn{2}{c}{2.7}& W\,m$^{-1}$\,K$^{-1}$ & \cite{y1981,ha1986} \\
Thermal conductivity of water & $k_l$ & \multicolumn{2}{c}{0.6}& W\,m$^{-1}$\,K$^{-1}$ & \cite{rnnnaw1995} \\
Kinematic viscosity of water at 0$^\circ$C & $\nu$ & \multicolumn{2}{c}{$1.8 \times 10^{-6}$} & m$^{2}$\,s$^{-1}$ & \cite{ofp1977} \\
\end{tabular}
\end{ruledtabular}
\end{table*}
Here we consider the melting of a sphere of mass $M_g = \rho_m 4/3\pi a^3$ at an initial temperature $T_s$ in a mass $M_l$ of well mixed, warmer and pure water (or at least, far from salt saturation) at temperature $T$, enclosed in a tank. The differential velocity between the sphere and the fluid is $U$. The total mass of the system is $M_{tot} = M_l + M_g$. Then, the energy conservation in the fluid -- including the liquid that has come from the melt and is still at the melting temperature $T_0$ -- can be written as
\begin{eqnarray}
C_{p}^{l} (M_{tot}-M_g) \frac{\partial T}{\partial t}&=&-4\pi a^2 F_T+\rho_0 C_p^l(T-T_0)4\pi a^2\frac{\partial a}{\partial t}+\mathcal{P}(T_{ext}-T)C_{p}^l (M_{tot}-M_g)
\label{eq:heat}
\end{eqnarray}
where $C_{p}^{l}$ is the specific heat capacity of water (we neglect its dependence on temperature and salt concentration) and $\rho_0$ the density of the $25\%$-NaCl water used to mold the sphere (we neglect its dependence on temperature). The first term of the right-hand side corresponds to the convective heat flux $F_T$ from the liquid towards the sphere. The second term corresponds to the heating of the liquid melt from its release temperature $T_0$ to the liquid temperature $T$ (we neglect the density change between the solid and the melt, and we assume rapid and complete mixing). The last term of the right-hand side corresponds to the heat losses by the liquid in the surrounding environment through the tank boundaries. $\mathcal{P}$ is an effective heat exchange or heat loss parameter, which can be experimentally determined by simply measuring the cooling of pure water in the same set-up but with no melting sphere. Note that it is negligible in both tanks for our main experiment on falling, melting spheres, but it has to be accounted for in our melting model validation experiment presented in the next section.

The salt mass conservation is given by 
\begin{equation}
M_{tot} c=M_l c_l+M_g c_0,
\end{equation}
where $c_l$ is the liquid concentration of sodium chloride ($c_l(t=0)=0$) and $c$ the mean concentration. The salt concentration in the sphere $c_0$, the mean concentration $c$, and the total mass $M_{tot} = M_l + M_g$ are all constant. Hence differentiating this equation gives 
\begin{equation}
\frac{\partial c_l}{\partial t} = \frac{(c_l - c_o)}{(M_{tot}-M_g)}\frac{\partial M_g}{\partial t}=\frac{(c_l - c_o)}{(M_{tot}-M_g)}\rho_m 4 \pi a^2\frac{\partial a}{\partial t}.
\label{eq:massrate}
\end{equation}

The mass of sodium chloride in our reactive sphere is small compared to the total volume of water. Then, the mean liquid concentration is small and far from the saturation point, hence the salt concentration in the liquid does not prevent the melting of the sphere. 

Following the Stefan condition at the melting interface, the growth rate is defined by a balance between the latent heat release due to the melting, the heat flux from the sphere towards the liquid, and the heat flux from the liquid towards the sphere
\begin{equation}
\left[L_{ice} \rho_{ice} (1-\Phi_{hh}) +L_{hh} \rho_{hh} \Phi_{hh} +L_{NaCl} \rho_{hh} \Phi_{hh} c_{pe}\right]\frac{\partial a}{\partial t}=k_m\left.\frac{\partial T}{\partial r}\right|_{r=a-} -k_l\left.\frac{\partial T}{\partial r}\right|_{r=a+}
\end{equation}
where $k_m$ and $k_l$ are the heat conductivity of the sphere and of the liquid, respectively. Here we take into account the binary mixture of the sphere, hence the release of latent heat due to ice melting $L_{ice}$ as well as the enthalpy due to hydrohalite dissociation $L_{hh}$ followed by sodium chloride dissolution $L_{NaCl}$. All processes are endothermic, which means that the melting of the sphere will absorb energy from the surrounding liquid. By assuming as a first order approximation a linear temperature profile through the sphere, we can write
\begin{equation}
\left[L_{ice} \rho_{ice} (1-\Phi_{hh}) +L_{hh} \rho_{hh} \Phi_{hh} +L_{NaCl} \rho_{hh} \Phi_{hh} c_{pe}\right]\frac{\partial a}{\partial t}=k_m \left(\frac{T_0-T_s}{a}\right) -F_T,
\label{eq:stefan}
\end{equation}
where $T_0$ is the temperature at the surface of the sphere equal to the melting temperature of ice, $T_s$ the temperature at the sphere center assumed to remain at its initial value, and $F_T$ the convective heat flux defined in Eq.~(\ref{eq:nusselt}). 

The convective heat flux $F_T$ has been measured for a large range of Reynolds, Prandlt and Schmidt numbers, as shown by Clift et al.~\cite{cgw1978}. We use here the parameterization given by Zhang and Xu~\cite{zx2003}, valid over a large range of Reynolds number
\begin{equation}
F_T =\frac{k_l(T-T_0)}{2a}  \left( 1 + (1+Pe_T)^{1/3}\left(1+\frac{0.096Re^{1/3}}{1+7Re^{-2}}\right) \right) 
\label{eq:nusselt}
\end{equation}
where $Pe_T = \frac{2 a U}{\kappa_l} $ is the Péclet number and $\kappa_l=\frac{k_l}{\rho_w C_p^l}$ the liquid thermal diffusivity. 

We can finally model the melting of a sphere by solving the two coupled differential equations (\ref{eq:heat}) and (\ref{eq:stefan}) in terms of the liquid temperature $T$ and the sphere radius $a$, using the physical properties from the table~\ref{tab_values} and the heat flux parameterization (\ref{eq:nusselt}). Equation (\ref{eq:massrate}) then gives the liquid concentration evolution $c_l$, and a polynomial fit of the equation of state of an aqueous sodium chloride solution (see Appendix~\ref{sec:appendix:melting}) finally allows evaluating the fluid density.

\subsection{Melting of a reactive sphere in a beaker}
\label{sec:results:melting}

\begin{figure}
\includegraphics[]{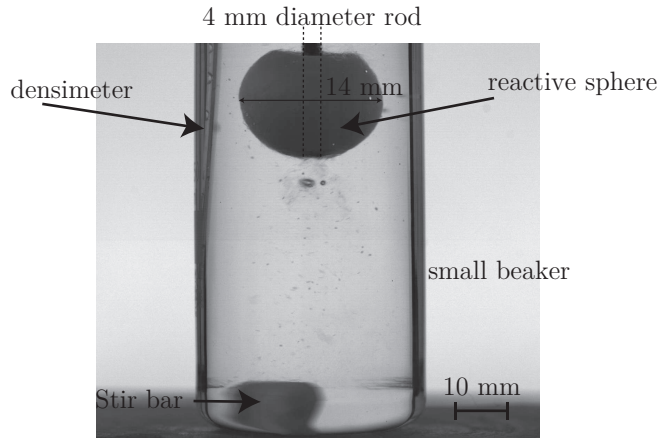}
\caption{Experiment in a small beaker of a sphere melting in a turbulent fluid using a magnetic stirrer.}
\label{fig:smallbeaker}
\end{figure}

\begin{figure}
\includegraphics[]{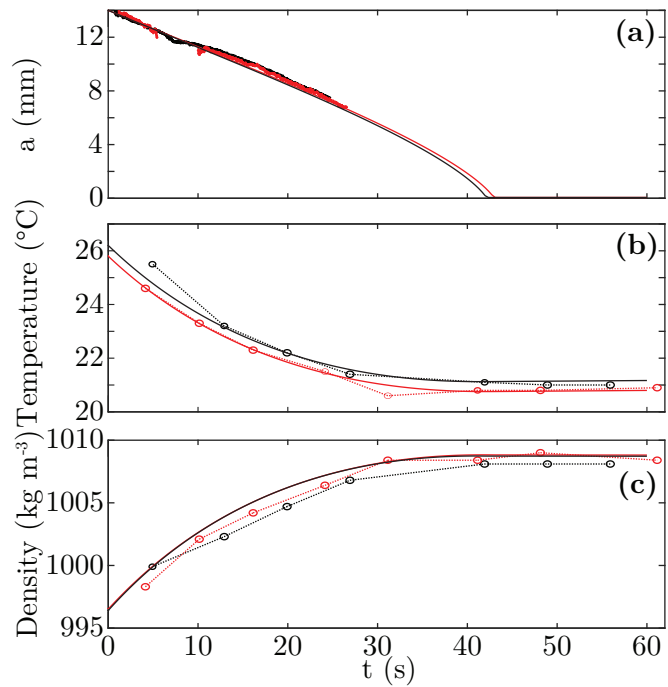}
\caption{Evolution of the radius (a), fluid temperature (b), and fluid density (c) for two similar experiments (red and black) of a reactive sphere melting in well-mixed pure water. Thick solid lines (in (a)) and dotted lines with circles (in (b) and (c)) denote measurements. Thin solid lines correspond to the model presented in Section~\ref{subsec:meltmodel} with a kinematic viscosity $ 9 \times 10^{-7}$~m$^{2}$\,s$^{-1}$ and a velocity $35$~cm\,s$^{-1}$.}
\label{becher}
\end{figure}

In this section, our goal is to validate the theoretical model of melting presented above. It would be very difficult to track the temperature and the salt concentration in the wake of a sphere during its fall in our complete experiment. Besides, it will turn out to be also very difficult to measure the radius changes, as will be discussed in the next paragraph. 
Thus, we have carried out a simpler experiment in a small beaker, where a sphere hanging from a rod is completely immersed in a known volume of pure water. The liquid is stirred with a magnetic stirrer to homogenize the temperature and the concentration during the melting. We performed density and temperature measurements in the fluid with a portable density meter (Anton-Paar DMA 35). Using a high-speed camera, we also tracked the radius evolution with time. We have measured previously the effective heat exchange / heat loss parameter of this set-up, $\mathcal{P} = 4.8 \pm 0.2 \times 10^{-4}$~s$^{-1}$. Fig.~\ref{becher} shows the evolution of the radius, temperature, and density for two runs with a 14~mm reactive sphere. The sphere falls from the rod after 20 seconds which prevents tracking the radius afterward. The typical fluid velocity has been estimated experimentally to be $\mathcal{O}(0.35)$~m\,s$^{-1}$. We solve the set of equations (\ref{eq:heat}), (\ref{eq:massrate}) and (\ref{eq:stefan}) with $a(t=0) = 14$~mm and $c_l(t=0)=0$~wt\%. Note that we do not take into account the effect of the rod, since it is small and heat transfer in metal is very rapid. The initial fluid temperature is the only free parameter as it has not been measured precisely. To solve Eq.~(\ref{eq:nusselt}), we have to estimate the value of the kinematic viscosity of the fluid. Since the Prandtl and Schmidt numbers are large, the thermal and chemical boundary layers are small compared to the viscous boundary layer. Besides, changes in the bulk fluid composition and temperature are small. Therefore, we consider the viscosity of the bulk of the fluid that is $\nu(\sim 25^\circ\text{C},\sim 0~\text{wt\%}) = 9 \times 10^{-7}$~m$^{2}$\,s$^{-1}$ (see Appendix~\ref{sec:appendix:melting}; temporal changes in temperature and concentration over the course of the experiment have no significant effect on viscosity). By adjusting, using a least squared method, the initial temperature to $T(t=0\,\text{s})= 25.8^\circ\text{C and } 26.2^\circ\text{C}$~respectively for the two considered experiments (in red and black in Fig.~\ref{becher}), the theoretical evolution of the radius, liquid temperature and density is in good agreement with measurements. 

\begin{figure}
\includegraphics[]{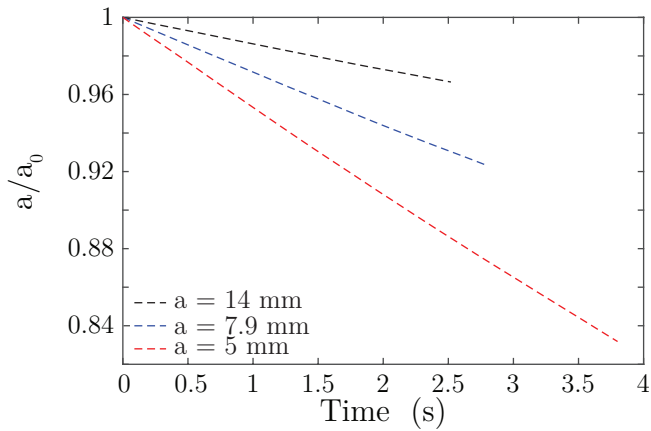}
\caption{Evolution of the radius normalized by its initial value (5 (red), 7.9 (blue), 14 mm (black) respectively) during the fall with a kinematic viscosity of $ 9 \times 10^{-7}$~m$^{2}$\,s$^{-1}$ (dotted lines). The falling velocity and time are based on an experiment in the large tank with a stratification $N = 0.164$~Hz (see Fig.~\ref{vel_strati_melt}).}
\label{fig:meltingrate}
\end{figure}

Using the same set of equations, we then estimate the melting rate of a sphere falling in a stratified layer in our complete experiment. Because the reactive spheres fall in a few seconds only, we consider that the temperature and the concentration of the far-field remain constant. Thus, our estimation of the melting rate is an upper bound: the increase of salt concentration and the decrease of the temperature in the fluid surrounding the melting sphere would indeed reduce the melting rate if taken into account. Here, we use the falling velocity (as detailed below in Fig.~\ref{vel_strati_melt}) for each different sphere radius as the value of $U$ in Eq.~(\ref{eq:nusselt}). Over the falling time in the stratified layer, the sphere radius decreases by less than, 2\%, 6\%, and 15\% for the 14, 7.9, 5 mm spheres respectively (see Fig.~\ref{fig:meltingrate}). With the high-speed camera equipped with a zoom ((2) in Fig.~\ref{fig:schema}), we observe qualitatively the melting of the reactive sphere. However, we observe the same variation of radius for both the plastic and reactive spheres. Indeed, the small radius variation induced by melting is of the same order of magnitude as the apparent decrease of radius due to refraction index changes with salt content. In the following, we thus use a constant radius for the reactive spheres to calculate all dimensionless parameters, including the drag coefficient, the Froude number, and the Reynolds number: this hypothesis only implies a small error which will be further discussed below in Fig.~\ref{fig:testdrag}.

\section{Dynamics of the plastic and reactive spheres}
\label{sec:fallingdrag}

\subsection{Falling behavior of the plastic and reactive spheres}
\label{subsec:fall}
\begin{figure*}
\includegraphics[]{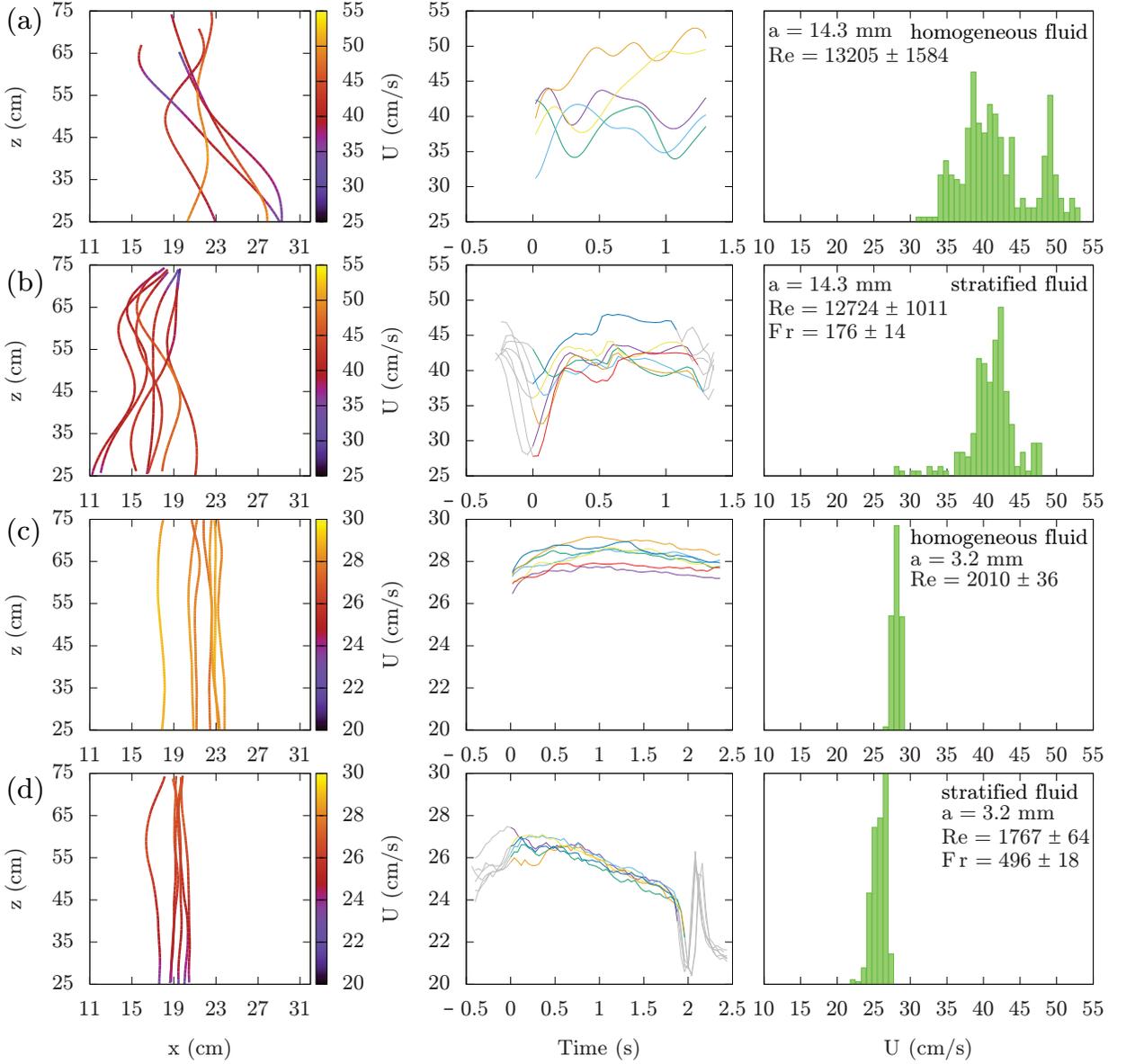}
\caption{Evolution of the velocity during the fall of the plastic spheres for 2 different radii (14.3 (a,b) and 3.2 mm (c,d) respectively) in a homogeneous fluid (a,c) and in a linear stratified layer with $N = 0.164$~Hz (b,d). The left column shows the horizontal and vertical position of each sphere, together with the norm of the velocity $U(z)=\sqrt{u_z^2+2u_x^2}$ ($u_z$ and $u_x$ are the vertical velocity and the horizontal velocity respectively) as a color scale. Note that in (b,d) the stratification starts just above the top of the figure and ends at the bottom of the figure, i.e. at $z=25~cm$. The middle column represents the evolution of {the norm of} the velocity {$U(z)$} for each sphere. The lines are colorized when the spheres are in the stratified layer, starting from $t=0$ at the top interface. The right column shows the distribution of {the norm of} the velocity {$U(z)$} of all spheres through the fall. The Reynolds number $Re$ and the Froude number $Fr$ are calculated with the median of the velocity distribution.}
\label{vel_strati}
\end{figure*}

\begin{figure*}
\includegraphics[]{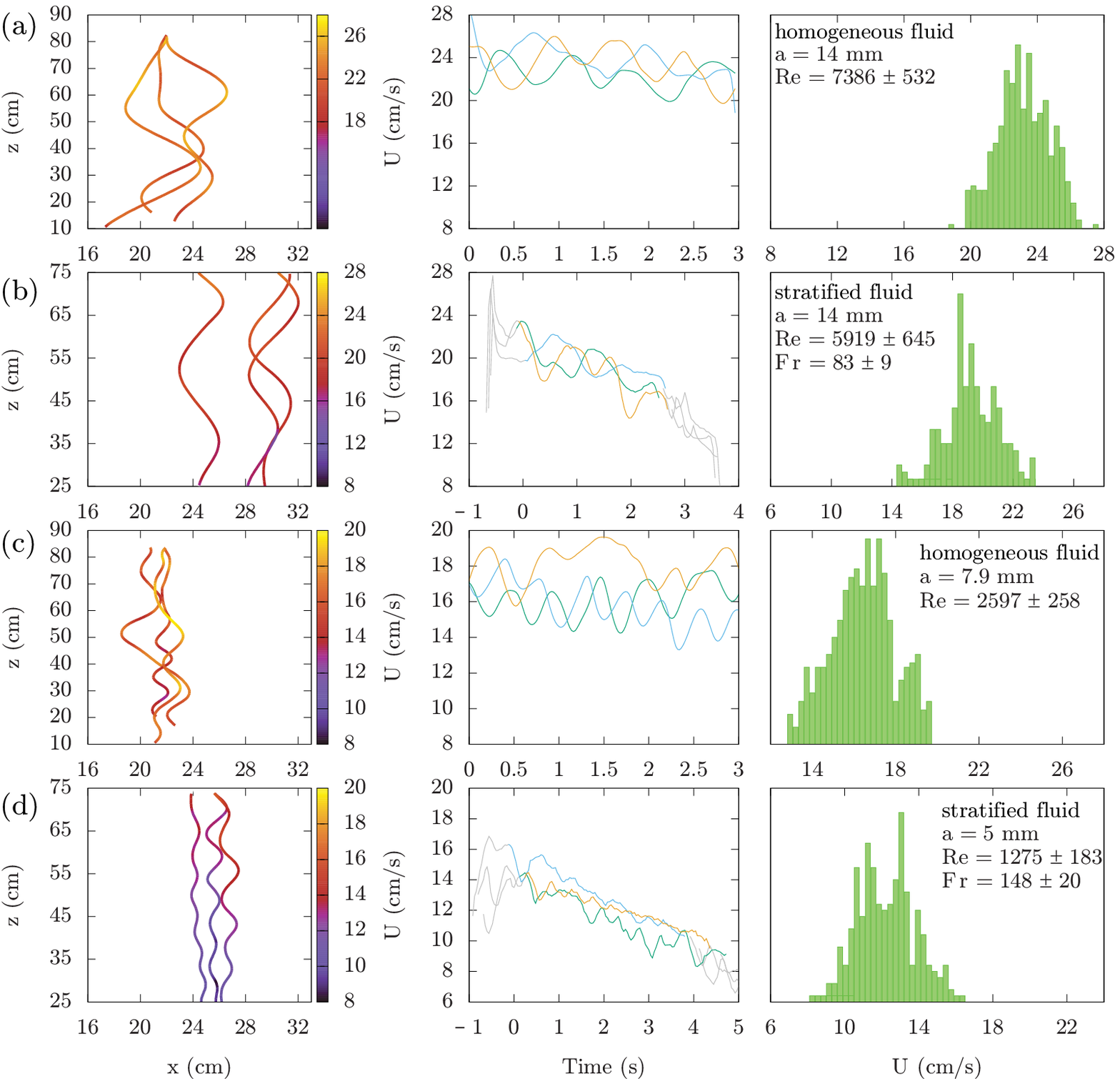}
\caption{Same as Fig.~\ref{vel_strati} for the reactive spheres (14 (a,b), 7.9 (c) and 5 mm (d) respectively) in the large tank with a homogeneous fluid (a,c) and with a linear stratification of $N = 0.164$~Hz (b,d). Note that (c) is for a different radius than (d).}
\label{vel_strati_melt}
\end{figure*}

Here, using a back-lighting technique, we present the results on the settling of plastic and reactive spheres. In all cases over the explored parameter range, the vertical settling is combined with oscillatory horizontal motions, with various but always small amplitudes (the ratio of horizontal vs. vertical velocities is always less than $10 \%$ and is becoming even smaller for the smallest spheres). We first performed two experiments in a homogeneous fluid, respectively in the small tank and in the large tank, to investigate the wall effects. Second, we conducted three experiments with a stably stratified layer (two in the small tank and one in the large tank) for three different Br\"unt-V\"ais\"al\"a frequencies $N=$ 0.136~Hz, 0.164~Hz and 0.196~Hz (see linear stratifications in Fig.~\ref{fig:strati}). Up to 7 releases at the top of the tank were performed for each sphere size and type. A low-pass filter was applied on the positions $x$ and $z$ to remove all high-frequency noise due to the post-processing, before calculating the velocity by finite-difference. The Reynolds number $Re$ and the Froude number $Fr$ are calculated with the median of the velocity distribution for each sphere size, the initial sphere radius and with the viscosity of the bulk fluid for the mean concentration along the density profile and at ambient temperature. Figs.~\ref{vel_strati}-\ref{vel_strati_melt} show the settling evolution of the plastic and reactive spheres in the larger tank with a stratification band (see Fig.~\ref{fig:strati}b), and in pure water. 

Fig.~\ref{vel_strati}b,d show the velocity of the plastic spheres for 2 different radii (14.3 and 3.2 mm), with time $t=0$ corresponding to the time when the sphere reaches the top of the stratified layer. Fig.~\ref{vel_strati}a,c show the same cases in a homogeneous fluid, for reference. Before entering the stratified layer, the velocity of the largest spheres rapidly decreases, and then increases once in the layer. This rapid variation is not observed for the smallest plastic spheres (3.2 and 1.6 mm). For all radii, however, a jump of velocity occurs when the spheres reach the bottom of the stratified layer. These observations agree with the recent study of Verso et al. \cite{vvl2019}, which relates velocity changes to the crossing of a relatively sharp density interface. In our experiment, the transition is always sharp at the bottom of the stratified layer (see Fig.~\ref{fig:strati}b), whereas it is seen as sharp at the top interface by large spheres only. 
Then, within the stratified layer (colored lines in the middle column of Fig. \ref{vel_strati}b,d), the velocity decreases with depth for the smaller spheres whereas for the larger spheres it remains almost constant. Besides, the falling path is quite periodic for the largest spheres, while the smallest spheres (3.2 and 1.6 mm radius) fall only with small oscillations. Actually, the velocity decrease is correlated to the relative decrease of the sphere buoyancy due to the ambient stratification: the density difference between the sphere and the fluid driving the fall decreases with depth because fluid density increases with depth. For the largest plastic spheres, this evolution is of the order of 6\% using the Newtonian velocity scaling introduced below (see (\ref{eq:generalstokes})), hence it is of the same order of magnitude as the variation due to the non-rectilinear motion: this is why the velocity decrease is not observed in these specific cases. The median velocities of the two largest spheres are almost similar (41.1 cm\,s$^{-1}$ and 40.7 cm\,s$^{-1}$ for the 14.3 mm and 7.9 mm spheres respectively), which is also likely due to the strong oscillations of the sphere. 
For comparison, we also show the trajectory of our plastic spheres in a homogeneous fluid for the same radii. For the largest spheres (top line in Fig.~\ref{vel_strati}a), the falling trajectories are more chaotic and the velocity spreads on a larger range. For the 3.2~mm radius spheres with a smaller Reynolds number (Fig.~\ref{vel_strati}c), the falling paths are quite similar to the stratified case, yet with a constant sinking velocity. As shown in \cite{hw2010}, the falling path is strongly perturbed by the noise background in the fluid, which we did not carefully remove between each launch. Nevertheless, for a stratified layer, we may expect that the motion noise is smaller, mostly two-dimensional, and more rapidly attenuated: therefore, the presence of a stratified layer extends the regime of regular oscillatory path.

Fig.~\ref{vel_strati_melt}b,d shows the velocity of two reactive spheres in a stratified layer, and Fig. \ref{vel_strati_melt}a,c in a homogeneous fluid for comparison. All reactive spheres oscillate rather strongly when settling into both stratified and homogeneous layers. The associated wavelength decreases with the size of the spheres ($\lambda \sim 30, 18, 10$~cm for radius $a = 1.43, 0.79, 0.5$~cm respectively). The ratio $\lambda/(2a)$ is about 10, which is close to the typical wavelength $\lambda/(2a)=12$ found for spheres falling in a homogeneous fluid with high Reynolds number \cite{mj1964,hw2010}. The presence of stratification does not seem to strongly modify this wavelength in the explored range of (rather large) Froude number. However, the trajectories are more periodic (i.e. less chaotic) when the spheres fall in a stratified layer, as already observed for plastic spheres. Moreover, the falling path of the reactive spheres seems to be even more regular than one of the plastic spheres, which may be due to smaller Reynolds number and density ratio $m^\star\sim 1.1$. These trajectory changes are also associated with amplitude oscillations of the velocity, which becomes smaller with a smaller sphere radius. With melting spheres in a stratified layer, we observe the expected velocity decrease with depth for all radii due to the buoyancy decrease: estimate based on the Newtonian velocity given below (see (\ref{eq:generalstokes})) indeed leads to a relative decrease of 14\% in velocity from the top to the bottom of the stratification. Note however that here, we do not see any rapid change around neither the top nor the bottom interface. All those unexpected dynamics would deserve a more detailed study with a dedicated set-up; here, we will simply focus on the effect of melting on the mean falling velocity.

\subsection{Drag coefficient of the plastic and reactive spheres}
\label{subsec:drag}
At large Reynolds number, the sedimenting velocity of a buoyant, solid sphere in a homogeneous fluid writes \cite{zx2003}
\begin{equation}
U=\sqrt{\frac{8ga\Delta \rho}{3\rho C_d^H}},
\label{eq:generalstokes} 
\end{equation}
where the drag coefficient $C_d^H$ scales as \cite{cgw1978}
\begin{equation}
C_d^H=\frac{24}{Re}+(1.8Re^{-0.313})+\frac{0.42}{1+42500Re^{-1.16}}.
\label{eq:cd}
\end{equation}
This parametrization is valid for Reynolds number $Re$ up to $3 \times 10^5$ with an error of about 5\%.

Here, we define the experimental instantaneous drag coefficient by
\begin{equation}
C_d^m(z)=\frac{8g a (\rho_m-\rho(z))}{3\rho(z) U(z)^2}
\label{eq:dragm}
\end{equation}
where $U(z)=\sqrt{u_z^2+2u_x^2}$ is the measured velocity of the sphere as a function of its vertical position. $a$ is the initial radius of the sphere which is considered constant even for the reactive sphere knowing that the melting rate is small (see Sec.~\ref{sec:results:melting} and further discussion in Fig.~\ref{fig:testdrag} below). $\rho(z)$ is the density profile in the stratified layer (Fig.~\ref{fig:strati}).
As for our melting model (see Sec.~\ref{sec:results:melting}), we also calculate the Reynolds number $Re$ using the kinematic viscosity of the bulk fluid with $T = 25^\circ$C and the salt concentration of the ambient fluid; but we acknowledge that the reactive spheres release cold and salty fluid which has a viscosity about three times larger. Our Reynolds number should thus be considered as an upper bound (see further discussion in Fig.~\ref{fig:testdrag} below).
We have calculated the value of the drag coefficient $C_d^m$, Reynolds number $Re$, and Froude number $Fr$ for all $z$ positions in all experiments with the corresponding values of $U(z)$ and $\rho(z)$. We have then defined for each configuration the median values and the standard deviations. The range of explored median Froude and Reynolds numbers is shown in Fig.~\ref{fig:regimediagram}.

\begin{figure}
\includegraphics[]{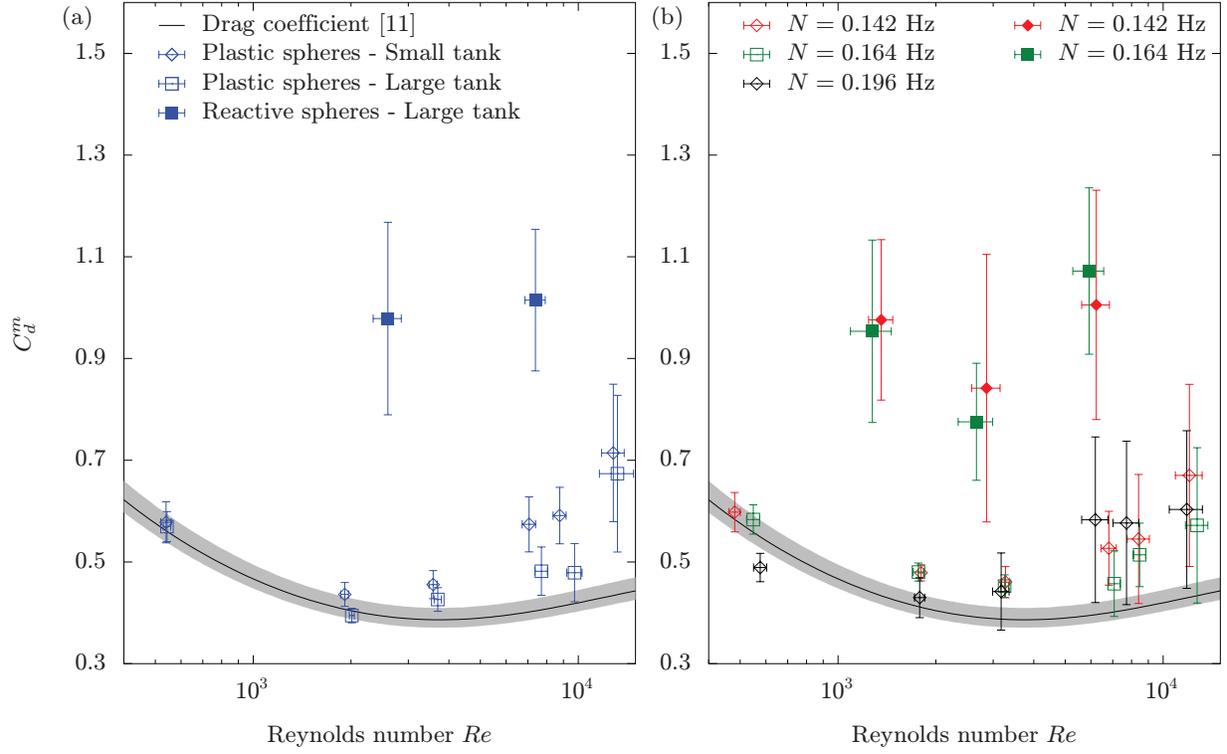}
\caption{Drag coefficient $C_d^m$ as a function of the Reynolds number $Re$. The solid line corresponds to Eq.~(\ref{eq:cd}) and the gray area denotes the +6\% and -4\% estimated uncertainty on this equation \cite{cgw1978}. The diamonds and squares denote the experiments in the small and large tanks, respectively, and the empty and filled symbols stand for the plastic and reactive spheres, respectively. (a) is for an homogeneous fluid and (b) for a stably stratified one.}
\label{fig:cdre}
\end{figure}

To test the significance of our choices regarding the sphere radius and the fluid viscosity, we have also calculated the drag coefficient with different hypotheses. First, we compare three different models of radius evolution: constant, model-based, and measure-based (with the limitations already described in section \ref{sec:results:melting}), shown respectively as green, blue, and red in Fig.~\ref{fig:testdrag}. A constant radius slightly overestimates the value of the drag coefficient and Reynolds number compared to the model-based and measure-based values. Then, considering the upper bound of fluid viscosity, i.e. the viscosity of the released fluid at the interface with $T = 0^\circ$C and $c=25$~wt\% (maroon dots in Fig.~\ref{fig:testdrag}), the Reynolds number is about three times smaller for the same drag coefficient. In all cases, however, a strong signature of melting on the drag coefficient is clearly observed compared to the non-reactive case in a homogeneous layer (Fig.~\ref{fig:cdre}a) or in a stratified layer (Fig.~\ref{fig:cdre}b).
\begin{figure}
\includegraphics[]{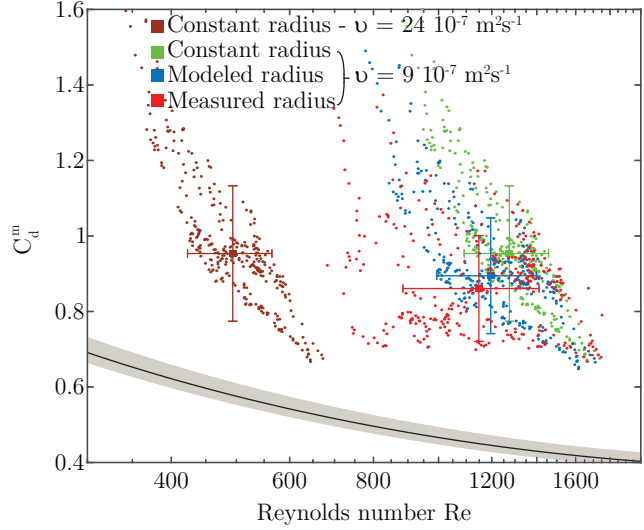}
\caption{Drag coefficient $C_d^m$ as a function the Reynolds number $Re$ for three 5 mm-radius reactive spheres falling in a stratified layer ($N=0.164$~Hz), considering various hypotheses for the sphere radius and the fluid viscosity. The solid line corresponds to Eq.~(\ref{eq:cd}) and the gray area denotes the +6\% and -4\%  estimated uncertainty on this equation \cite{cgw1978}. Small dot markers denote instantaneous values during the fall of each sphere, while the large squares show corresponding median values and standard deviations.}
\label{fig:testdrag}
\end{figure}

Fig.~\ref{fig:cdre}a shows the median drag coefficient $C_d^m$ as a function of the Reynolds number $Re$ for all experiments without stratification. For $Re<4000$, the drag coefficient of plastic spheres is in very good agreement with the scaling law of Clift et al.~\cite{cgw1978} for both tanks, hence confirming the absence of significant wall effect. For larger $Re$, wall effects seem more important for the smaller tank, where measurements deviate from the theoretical model. However, even the largest spheres have a radius of at least 20 times smaller than the tank width, and those effects remain limited. In all cases, the observed change in the drag coefficient due to the melting of reactive spheres is significantly larger, with a $C_d^m$ at least two times larger than the one of non-reactive spheres for the same $Re$ (full blue squares in Fig.~\ref{fig:cdre}a).

\begin{figure*}
\includegraphics[]{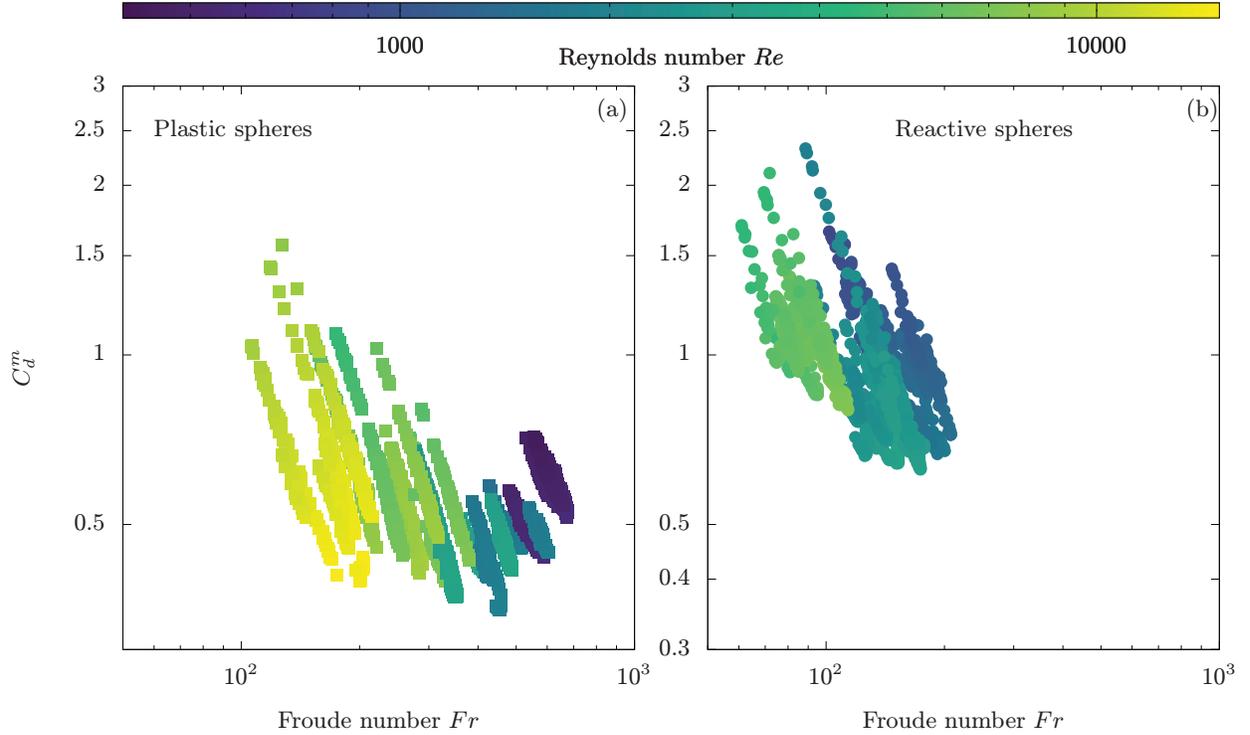}
\caption{Measured drag coefficient as a function of the Froude number $Fr$ at all $z$ positions for plastic spheres (a) and reactive spheres (b). The color scale denotes the Reynolds number $Re$. (For interpretation of the references to color in this figure legend, the reader is referred to the web version of this paper.}
\label{fig:cd_frre}
\end{figure*}

Fig.~\ref{fig:cdre}b additionally shows the drag coefficient for the experiments with the reactive and non-reactive spheres in a stratified layer. For all spheres, the drag coefficient is enhanced by stratification for all Reynolds numbers (see Fig. \ref{fig:cd_frre}). However, the effect of the stratification remains small because in our experiments the stratification is weak compared to the sphere velocity (i.e. high Froude number). Again, the effect of melting on the drag coefficient is predominant. 

The drag coefficient can also be altered by the shape and roughness of a falling sphere, and we acknowledge that our reactive spheres do not have a perfectly spherical shape due to their molding process. In Newton's regime (large Reynolds number), oblate spheroids ($b/a<1$ with $a$ and $b$ the major and minor axes respectively) have a larger drag coefficient than the one for the sphere. However, our spheres have a small flatness ($b/a\sim 0.9$) which implies a very small drag coefficient increase \cite{cgw1978,l2008}. Moreover, the roughness of a sphere only modifies the drag coefficient for $Re$ over $10^4$. One can notice that the density of the melt released at the surface of the reactive spheres is only slightly larger (10\%) than the density of the reactive sphere. Then, the added drag cannot be due to the added or lost mass during the melting because the total mass is almost conserved. In other words, the sphere and the fluid released at its surface fall together and the melting does not change the total buoyancy. Therefore, following Zhang et al., \citep{zmm2019}, we relate the strong drag enhancement of the reactive sphere to an increase of the vorticity induced by the released melt in the wake, where we indeed observe strong mixing.  The release of the melt at the sphere's surface induces a major density disturbance, which strongly interacts with the pressure field generated by the fall. Therefore, we expect the melt to drastically change the flow in the wake by inducing baroclinic torques that modify the vorticity field. This suggested mechanism now requires specific theoretical and numerical investigation, following the approach of \citep{zmm2019}.

\begin{figure*}
\includegraphics[]{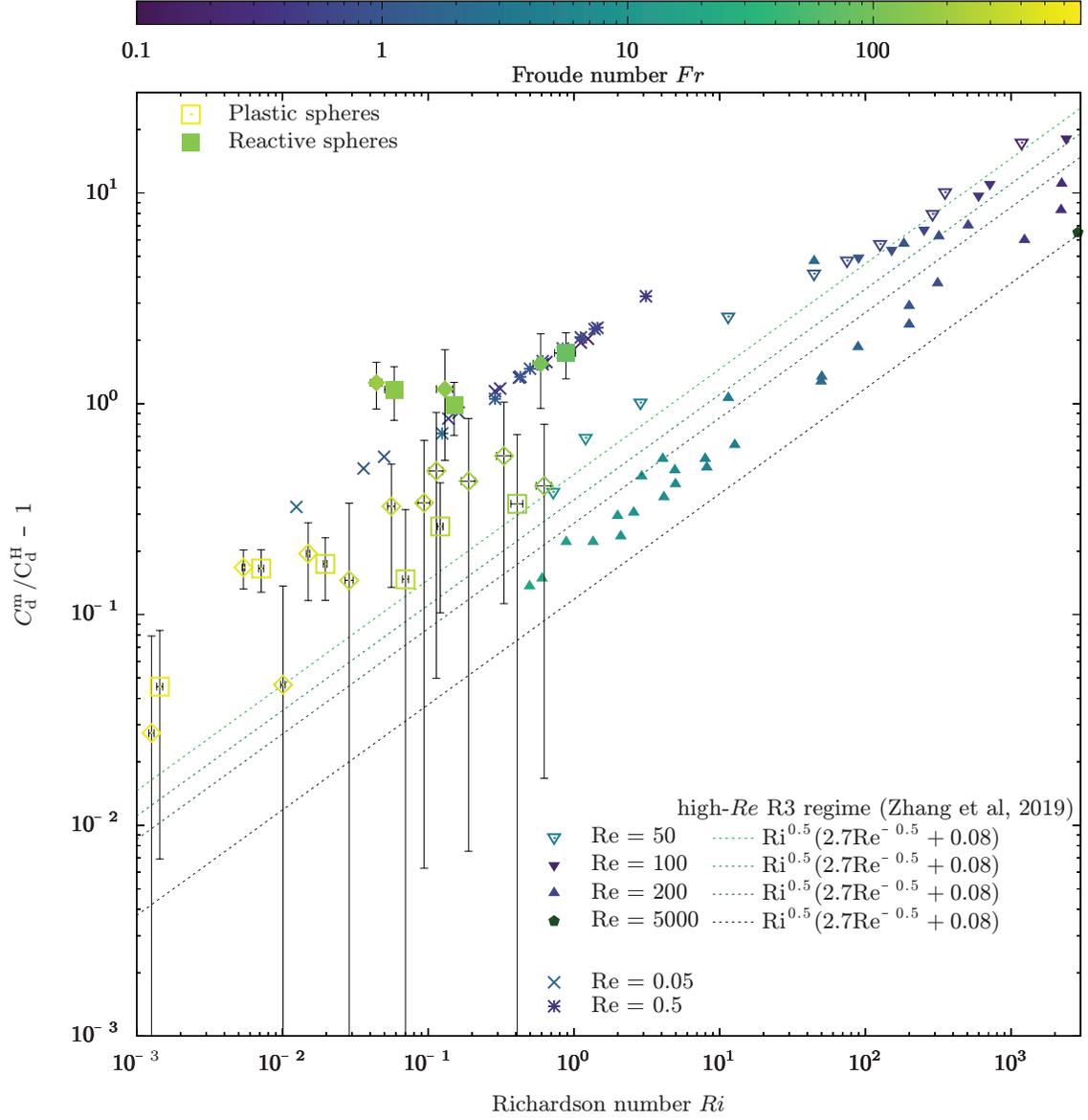}
\caption{Added drag coefficient $C_d^m/C_d^H-1$ as a function of the Richardson number $Ri$. Our experimental results are the squares and diamonds (filled or empty for reactive or plastic spheres respectively). The filled and empty triangles denote the previous numerical results of \cite{thocv2000,hko2009a,hny2015}. The pentagon denotes an Argo float \cite{d2003}. The crosses and stars correspond to the numerical results of Yick et al.~\cite{ytps2009}. The dashed lines represent the high Reynolds number regime suggested by Zhang et al.~\cite{zmm2019} for different values of $Re$. Here, all data are for a Schmidt number of 700. The value of the Froude number for each data point is scaled with the color bar. (For interpretation of the references to color in this figure legend, the reader is referred to the web version of this paper.)}
\label{fig:cd_Ri}
\end{figure*}

In the literature, the added drag coefficient due to the stratification $C_d^m/C_d^H-1$ is often described as a function of the Richardson number \cite{thocv2000,hko2009a,hny2015,ytps2009}, where the Richardson number is defined following Yick et al.~\cite{ytps2009} as
\begin{equation}
Ri = \frac{Re}{Fr^2}=\frac{2a^3 N^2}{\nu U},
\end{equation} 
comparing the buoyancy forces to the viscous shear forces. Our results are shown in Fig.~\ref{fig:cd_Ri}, together with the scaling laws found by Zhang et al.~\cite{zmm2019} and some previous numerical results \cite{thocv2000,hko2009a,ytps2009,hny2015}. Note that the numerical results of Yick et al. \cite{ytps2009} (stars and crosses in Fig.~\ref{fig:cd_Ri}) explore the same range of Richardson number but have a much smaller Froude number. For both reactive and plastic spheres, the added drag increases with the Richardson number, the added drag being about two times larger for the reactive spheres than for the plastic ones at a given Richardson number. Using a balance between the Archimedes, inertial and vorticity forces, Zhang et al.~\cite{zmm2019} predicted a scaling law for the added drag $C^m_d/C^H_d-1$ as $Ri^{0.5}(2.7Re^{-0.5}+0.08)$ for high Schmidt and Reynolds numbers, but moderate Froude number. Whereas this scaling is in a quite good agreement with the numerical results for $Re = 50,100,200$ of \cite{thocv2000,hko2009a,hny2015} and the Argo float value \cite{d2003}, it is not able to quantitatively explain our results for either plastic or reactive spheres. Our experimental results on the non-reactive plastic spheres might still scale with $Ri^{0.5}$, but a new study following the formalism of Zhang et al.~\cite{zmm2019} for high Froude numbers is necessary to infer the relevant prefactor. 
Besides, the added drag for the reactive sphere seems to show a shallower Richardson number dependence, even if this will require additional experiments for confirmation.

In conclusion, despite exploring a limited range in sphere radius and stratification, and despite intrinsic experimental limitations, our experiments exhibit a significative drag enhancement due to the melting and a specific dependence on stratification that will deserve additional studies, especially 3D numerical simulations following the recent results of Zhang et al.~\cite{zmm2019}.

\section{Internal waves after the fall of a sphere}
\label{sec:waves}
\subsection{Linear theory for plane waves with viscosity}
The dispersion relation of a linear internal wave in a viscous fluid writes
\begin{equation}
\left(\frac{\omega}{N}\right)^2-\textit{i}\nu \frac{k^2}{N}\left(\frac{\omega}{N}\right)-\frac{k_x^2+k_y^2}{k^2}=0,
\end{equation}
where $k$ is the wave-number with $k^2=k_x^2+k_y^2+k_z^2$, and $\omega$ the wave frequency. Waves with frequency $\omega > N$ are evanescent. For lower frequency waves, the roots of this second order polynomial can be split in a real part and an imaginary part as
\begin{equation}
\omega_r = \pm N \sqrt{\frac{k_x^2+k_y^2}{k^2}-\frac{\nu^2k^4}{4N^2}}\hspace{0.5cm}\text{and}\hspace{0.5cm}\omega_i = \frac{\nu k^2}{2}.
\end{equation}
The real part gives the oscillation frequency including the viscous shift; however, the viscous contribution is small when considering Br\"unt-V\"ais\"al\"a frequencies and wavelengths relevant for our experiments. The imaginary part corresponds to the wave damping for the given wavenumber.

The fall of each sphere creates turbulence in the wake, and emits a series of propagating waves of various frequencies and wave numbers. Hence the velocity field writes
\begin{equation}
\vec{u}=\sum \vec{u}_0 e^{\textit{i}(\vec{k} . \vec{x}-\omega_r t)-\omega_i t}
\end{equation}
where $\vec{u}_0=u_0\vec{e}_z+v_0\vec{e}_x+w_0\vec{e}_y$. In our experiments, the PIV measurements allow us to measure $u\vec{e}_z$ and $v\vec{e}_x$, and assuming isotropy in the horizontal plane, we estimate that $v\vec{e}_x \simeq w\vec{e}_y$.
Then, we can write the kinetic energy in the whole tank as
\begin{equation}
E_k=\frac{1}{2}\int_V \rho(z) \left(|u|^2+2|v|^2\right) dV. 
\label{eq:kin}
\end{equation}
Focusing on the least damped component with the smallest wave-number $k$ (see discussion below), the long time evolution of our PIV measurements then gives a relevant estimate of its attenuation rate 
\begin{equation}
\omega_I^{E_k}=\nu k^2. 
\label{eq:attrate}
\end{equation}
Looking at the initial value $E_k(t=0)$ indicates the amount of energy deposited in the wave field.

After multiple reflections on the domain boundaries, waves might form standing modes, whose velocity is discretized to account for the boundary conditions, with $k_x=m\frac{\pi}{L}, k_y=l\frac{\pi}{L}, k_z=j\frac{\pi}{H}$,
with $L$ and $H$ the width and the height of the tank and $m,l,j$ integers. Assuming isotropy in the horizontal direction means $m=l$.

\subsection{Results}

Here we analyze the experiments carried in the small tank with non-reactive and reactive spheres and three different Br\"unt-V\"ais\"al\"a frequencies $N =$~0.094~Hz, 0.136~Hz, and 0.196~Hz (see Fig.~\ref{fig:strati}a). We perform long-term PIV measurements, i.e. over durations between 120 and 180~second after the launch of a sphere, using the same set of PIV parameters in all experiments (that are box size, overlap, and time step). Between 2 and 4 launches are done for each of the three sizes of reactive and plastic spheres.

\begin{figure}
\includegraphics[]{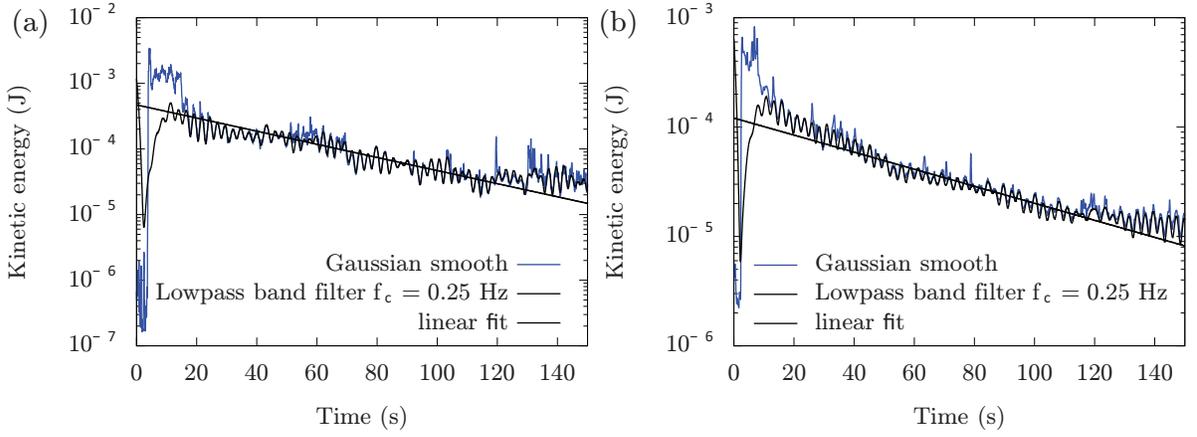}
\caption{Evolution of the kinetic energy in the stably stratified layer with $N =$~0.196~Hz when (a) a plastic sphere and (b) a reactive sphere of 14~mm-radius fall. The blue line is smoothed with a moving average to remove all high-frequency noise ($>2$\,Hz). The black line is filtered with a low-pass filter with a cut-off frequency $f_c= 0.25$~Hz to keep only the signature of propagating internal waves. An exponential fit is computed for the signal between 20~s and 130~s. (For interpretation of the references to color in this figure legend, the reader is referred to the web version of this paper.)}
\label{fig:energy1}
\end{figure}

In Fig.~\ref{fig:energy1}, we show the typical time evolution of the kinetic energy in the stably stratified layer for a plastic and a reactive sphere fall. The shown behavior is generic to all performed experiments. The energy drastically increases once the sphere is released in the tank and during its fast fall. Due to the large velocity of the fluid during these early times compared to the frame rate of our camera ($50$~FPS), the first $\sim$10 seconds of measurements should not be used quantitatively. However, the difference between the (blue) total energy and the (black) filtered energy over the wave propagating frequency domain shows that (i) at first, the falling of the sphere generates motion in all frequencies, presumably due to the turbulent wake, and (ii) after 20~s typically, all the energy is in the internal waves only. The kinetic energy then decreases close to exponentially due to the viscous dissipation of internal waves, while visibly oscillating at a frequency close to $2N$.

\begin{figure}
\includegraphics[]{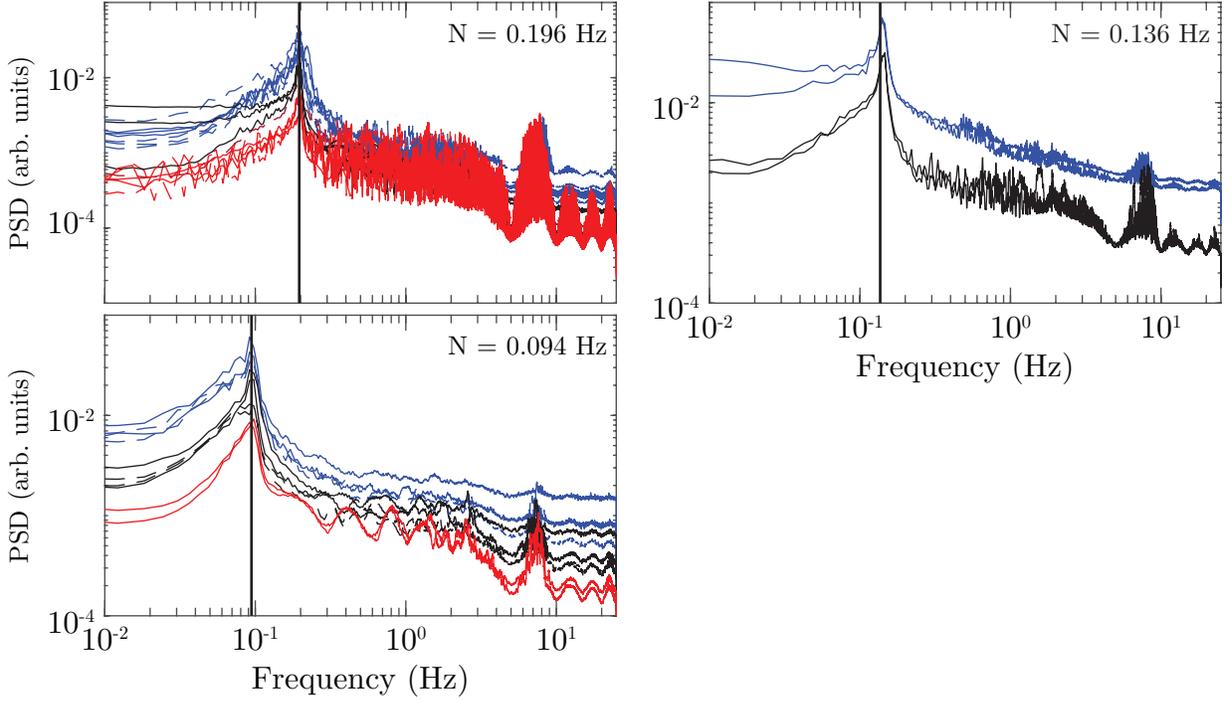}
\caption{Spatial mean of the frequency spectrum of the vertical velocity field for three different stratifications with Br\"unt-V\"ais\"al\"a frequency (vertical bold black line) $N =$~0.094~Hz, 0.136~Hz, and 0.196~Hz. Dashed and solid lines correspond to the reactive and plastic spheres respectively. Blue, black and red denote three different radius 14, 7.9 and 5 mm for the reactive spheres and 14.2, 7.9, 4.8 mm for the plastic spheres.}
\label{fig:ffttime}
\end{figure}

\begin{figure}
\includegraphics[]{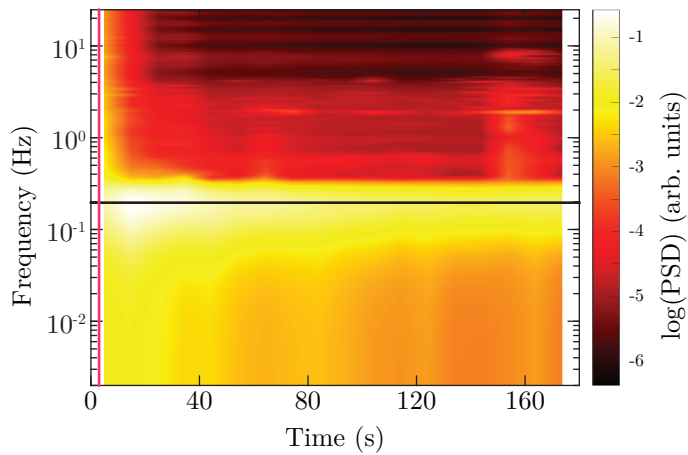}
\caption{Spatially averaged spectrogram of the vertical velocity in the stably stratified layer with $N =$~0.196~Hz (black line) during and after the fall of a 14~mm-radius reactive sphere. Spectrum energy is an arbitrary log-scale. The vertical red line denotes the launch time of the sphere.}
\label{fig:spectrogram}
\end{figure}

To further analyze this time dependency, Fig.~\ref{fig:ffttime} shows the frequency spectrum of the vertical velocity field for all the experiments. A marked peak is present at the Br\"unt-V\"ais\"al\"a frequency for all cases, with a strong cut-off for larger frequencies, corresponding to evanescent waves. 
This is further illustrated in Fig.~\ref{fig:spectrogram}, showing an example of the spectrogram for the vertical velocity field, spatially averaged over the whole domain. The fall of the sphere initially excites all frequencies. But rapidly, a cut-off appears above the Br\"unt-V\"ais\"al\"a frequency $N$, while frequencies smaller than $N$ are more progressively attenuated, the least damped component being at $N$.

\begin{figure}
\includegraphics[]{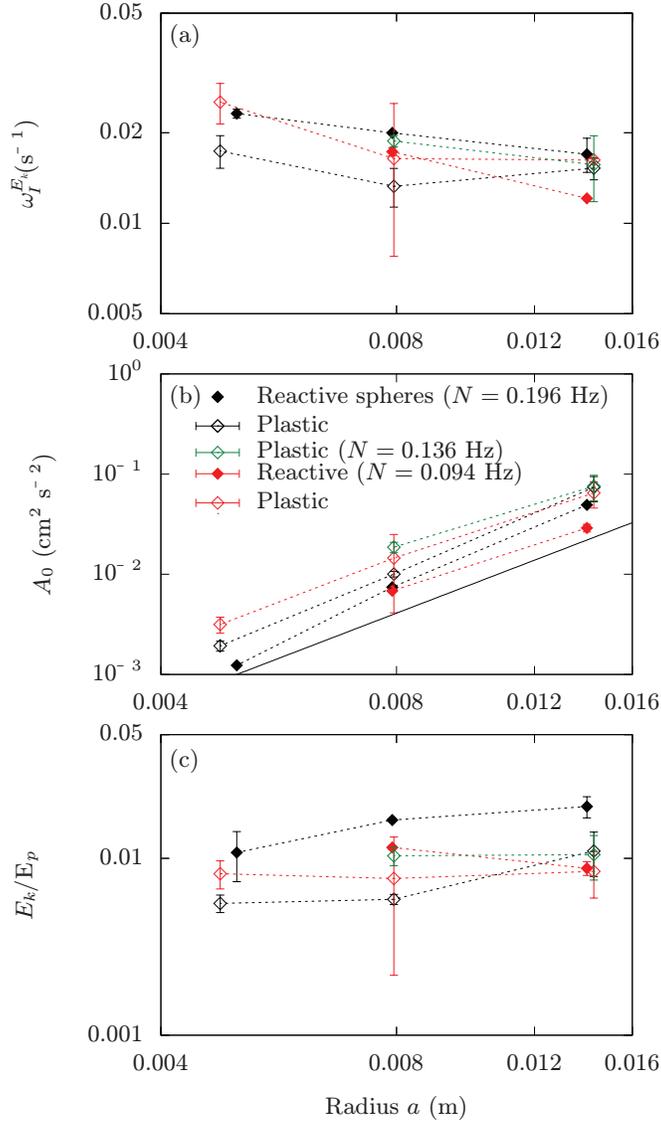}
\caption{The attenuation $\omega_I^{E_k}$ (a), the initial wave amplitude $A_0$ (b) and the ratio between the initial wave kinetic energy $E_k(t=0)$ and the potential energy $E_p$ (c) as a function of the radius of the spheres for three different stratifications with a Br\"unt-V\"ais\"al\"a frequency $N =$~0.094~Hz (red), 0.136~Hz (green), and 0.196~Hz (black). The empty and full diamonds denote the plastic and reactive spheres, respectively. The error bars denote the standard deviation based on several experiments performed with the same spheres (between 2 and 4 launches for each size and each stratification). In (b), the solid black line shows an $a^3$ slope. (For interpretation of the references to color in this figure legend, the reader is referred to the web version of this paper.)}
\label{fig:energy}
\end{figure}

For all our experiments, we calculate the initial amplitude and the attenuation rate of the least damped wave component of the energy signal by linearly fitting the log of the kinetic energy-filtered with a low-pass filter just above $N$ (see e.g. the black line in Fig.~\ref{fig:energy1}). Fig.~\ref{fig:energy} shows, as a function of the radius, the attenuation $\omega_I^{E_k}$ (top), the initial amplitude $A_0=2E_k(t=0)/\rho$ (middle), and the ratio between the initial wave kinetic energy $E_k(t=0)$ and the initial potential energy of the sphere at the top of the tank $E_p=\frac{4\pi}{3}a^3\Delta \rho gH$. The error bars denote the spread between different launches of similar spheres in each stratification. Fig.~\ref{fig:energy} does not exhibit any clear dependency on the Br\"unt-V\"ais\"al\"a frequency nor the sphere composition. The attenuation rate $\omega_I^{E_k}$ might slightly decrease with a radius for the plastic and reactive spheres, but this demands confirmation over a larger range. The initial amplitude $A_0$ increases with the radius, which seems reasonable since the energy injected into the system, i.e. the initial potential energy of the sphere, also increases with radius. A scaling law $A_0 \propto a^3$ seems to fit with our observations over the explored limited range (black line in Fig.~\ref{fig:energy}b): this implies a constant ratio $E_k(t=0)/E_p$, as indeed shown in the bottom figure. The part of potential energy dissipated in propagating waves is thus about 1\%: this ratio is quite similar to the amount of energy radiated from a turbulent mixed layer into a stratified layer \cite{ms2014} and from an impulsive plume \cite{bs2019}. But it is smaller than the energy dissipated from a buoyant parcel of fluid rising in a stratified layer \cite{c1978}.

\begin{figure}
\includegraphics[]{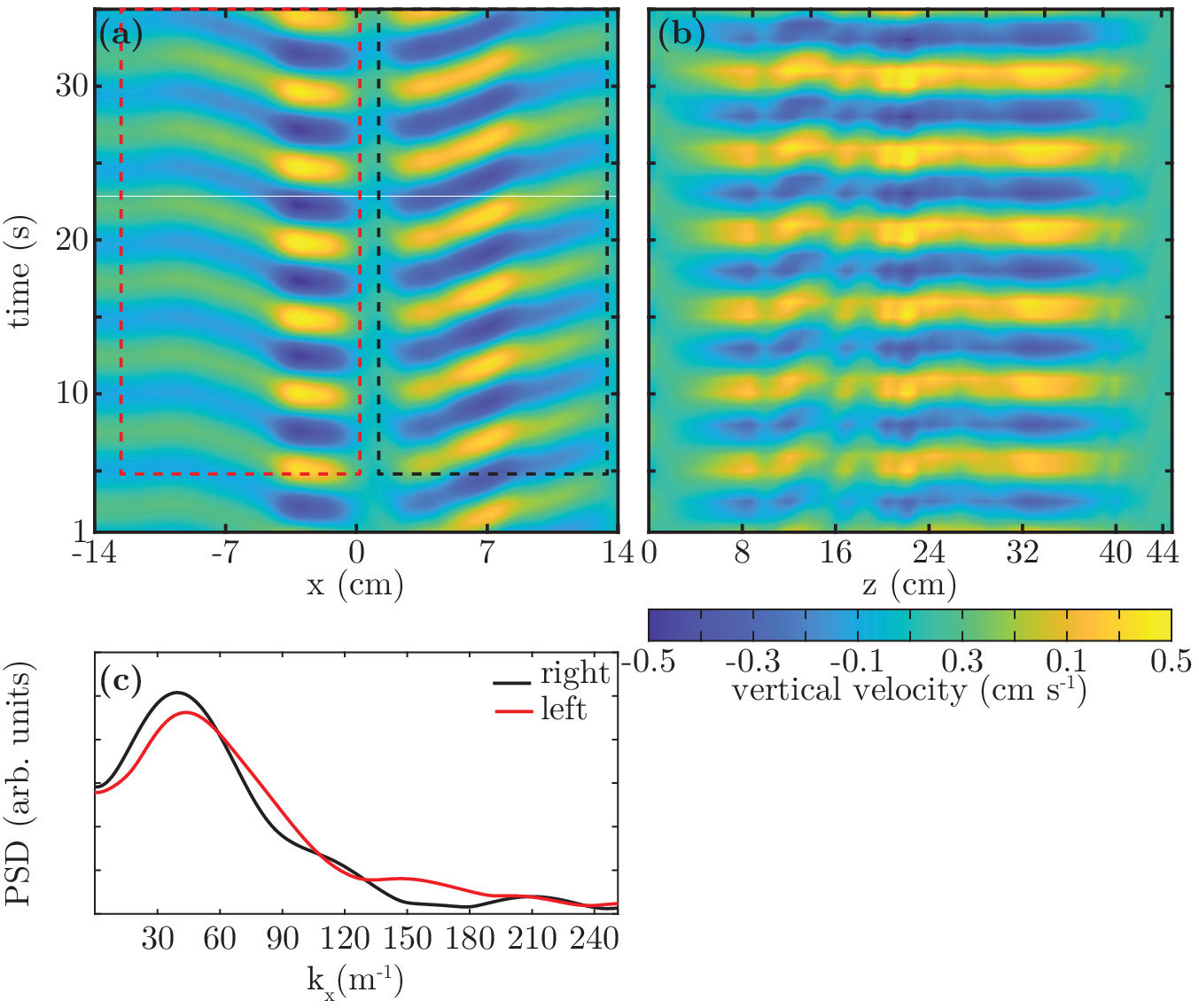}
\caption{(a) and (b) Spatiotemporal diagram of the vertical velocity induced by a plastic sphere (14.3 mm radius) falling in a stably stratified layer with $N=0.196$~Hz as a function of $x$ at mid-tank depth (a) and as a function of $z$ in the middle of the tank (b). The vertical velocity is filtered with a pass-band filter around $N$. (c) Power density spectrum for $k_x$ for the two radially symmetric windows highlighted in (a).}
\label{fig:kxkz}
\end{figure}

Those various experimental observations can be simply explained by noticing that the sphere falling time and the sphere radius are small compared to the buoyancy period and the tank dimensions, respectively: hence, from the internal wave point of view, at first order, the sphere fall might be considered as an impulsive excitation in time and the horizontal direction, and as an essentially uniform excitation in the vertical direction (with additional turbulent fluctuations). As such, it provides energy in all $\omega$ and $k_x$, and mostly in $k_z=0$. Following the dispersion relation, $k_z=0$ means $\omega = N$, and according to wave damping, the least damped component is at the smaller acceptable $k_x$. Since the sphere falls in the middle of the tank, the axial symmetry of the excitation imposes a minimum wavenumber $k_x=3\pi/L=31~\text{m}^{-1}$. This is indeed confirmed in Fig.~\ref{fig:kxkz}, showing the spatio-temporal diagrams of the vertical velocity during the first 30 seconds after a sphere fall. By measuring the wavenumber $k_x$ in two windows symmetric compared to the falling path, $k_x$ is about $39~\text{m}^{-1}$ (Fig.~\ref{fig:kxkz}a,c), while motions indeed seem independent of $z$ (Fig.~\ref{fig:kxkz}b).
Regarding dissipation, using those wavelengths with $\nu \sim 9\, 10^{-7}$~m$^2$\,s$^{-1}$ gives an attenuation rate $\omega_I^{E_k}=2.7\, 10^{-3}$~s$^{-1}$, which is one order of magnitude smaller than the results of Fig.~\ref{fig:energy}a.
Actually, $\omega_I^{E_k}$ measurements include the contribution of all waves as well as of the boundary dissipation, while our model only considers the bulk dissipation of the most long-standing wave. Finally, the actual geometry where the small sphere produces an initially axisymmetric perturbation that bounces out in a rectangular tank complexifies the oversimplified description provided above. The main observations are nevertheless in good agreement with our theory, including a single axially symmetric pattern in the horizontal direction, and the independence of the attenuation rate with the sphere radius.

\section{Conclusion}
\label{sec:discussion}

We have experimentally studied the fall of a sphere in a stably stratified layer, expanding previous works by exploring a regime of larger Reynolds and Froude numbers. We have focused our experiments on the specific behavior of a reactive (i.e. melting) sphere vs. an inert one. Besides, we have examined the fluid motions in our tank over several minutes after the rapid fall of a sphere (a few seconds) to investigate the generated internal wave field. As for previous studies at moderate Reynolds number, we show that the drag coefficient of a falling non-reactive sphere is enhanced due to the stratification and that the added drag might be proportional to $Ri^{0.5}$. However, the scaling law for moderate Reynolds and Froude numbers suggested by Zhang et al.~\cite{zmm2019} was unable to quantitatively predict our results: this is likely due to the more turbulent wake that induces both 3D motions and supplementary buoyancy contribution to the drag that are unaccounted for in previous numerical and theoretical studies.
The drag coefficient of melting spheres is strongly enhanced compared to non-reactive spheres. This enhancement is larger than any estimated shape and roughness effect. Besides, since the density of the released melt is close to the one of the melted sphere, the added drag cannot be due to an added mass associated with the melting. We think that it is actually due to the strong mixing in the wake of the sphere induced by melting, expending upon the recent results of \cite{zmm2019}. Finally, for both melting and inert spheres, an internal wave field encompassing about $1\%$ of the initial potential energy of the sphere is excited in the tank by the sphere fall. Since this fall is seen as an almost impulsive excitation in space and horizontal direction and an almost uniform excitation in the vertical direction, most of the released energy rapidly focuses at the Br\"unt-V\"ais\"al\"a frequency and at the largest admissible horizontal wavelength, where it is dissipated very slowly. 

We acknowledge that our study is limited by several factors, including the small explored range in terms of buoyancy frequency and sphere radius, as well as the slow melting rate of our sphere during their fall. Additional studies are necessary to investigate exhaustively the effect of melting on the sphere dynamics, expending upon the first conclusions drawn here. They should, in particular, focus on the motion in the wake behind the melting sphere. This is definitely an experimental challenge. Direct numerical simulations of a reactive sphere falling in a stratified layer, even for lower Reynolds number, would also be extremely useful to unravel the contribution of the release fluid at the melting surface and the modification of the wake. 

To finish with, let us simply mention the possible application of our results to planetary core dynamics. For small planets like Mercury and Ganymede, the top-down crystallization of their liquid iron core \cite{had2006,rbs2015} implies the formation of crystals at the top of the core and their sinking in a hotter region, hence partial or complete melting. The crystal size-range is poorly constrained but could vary from micrometer-scale \cite{rbs2015} to kilometer-scale \cite{hhvj2018,nbn2019}. A larger drag coefficient would imply a longer residence time of these falling crystals and locally modify the equilibrium state \cite{dp2018}. Even more interesting, the top of those planetary cores is often associated with a thermal or chemical stratification which prevents large-scale convective motions \cite{wh2014,rbs2015}, hence questioning the origin of their magnetic field. The sinking of large crystals could redistribute a small amount of their potential energy into kinetic energy via long-standing internal waves, which would help to provide the necessary kinetic energy to drive a long-lived dynamo.

\section{Acknowledgment}
This work was supported by the ERC (European Research Council) under the European Union's Horizon 2020 research and innovation program through Grant No. 681835-FLUDYCO-ERC-2015-CoG. We thank the two anonymous reviewers for their constructive reviews which help to improve our paper.

\newpage

\appendix

\section{Viscosity and density of an aqueous sodium chloride solution}
\label{sec:appendix:melting}

Following the density table of \cite{sgrg2015} for a salty water and the 5th-order polynomial standard equation for the pure water density \cite{k1975b,jh1992,mk2007}, the density of the aqueous sodium chloride solution as a function of temperature and composition at ambient pressure (1 bar) is given with a 5th-order polynomial fit
\begin{eqnarray}
\rho(c,T)=&b_1 + b_2T + b_3c + b_4T^2 + b_5Tc + b_6c^2 + b_7T^3\nonumber \\
&+ b_{8}T^2c + b_9Tc^2 + b_{10}c^3 + b_{11}T^4 + b_{12}T^3c\nonumber \\
&+ b_{13}T^2c^2 + b_{14}Tc^3 + b_{15}c^4 + b_{16}T^5 + b_{17}T^4c\nonumber \\
 & +b_{18}T^3c^2 + b_{19}T^2c^3 + b_{20}Tc^4 + b_{21}c^5
\label{eq:coeffrho}
\end{eqnarray}
with $T$ in Celsius and $c$ in wt\%. All coefficients are given in table \ref{montableau}.
\begin{table}
\caption[]{\label{table:coeffrho} Coefficient for Eq.~(\ref{eq:coeffrho})}\label{montableau}
\begin{ruledtabular}
\begin{tabular}{ll} 
$b_1$& 999.83952 \\
$b_{2}$& $4.58004182743 ~10^{-2}$ \\
$b_{3}$& $7.7259007462164 $ \\
$b_{4}$& $-81656026808 ~10^{-3}$ \\
$b_{5}$& $-2.89974283744 ~10^{-2}$\\
$b_{6}$& $-2.80856566638 ~10^{-2} $ \\
$b_{7}$& $6.39597015~10^{-5} $\\
$b_{8}$& $3.870476539 ~10^{-4}$ \\
$b_{9}$& $1.267276467~10^{-4}$\\
$b_{10}$ & $3.5362104886~10^{-3}$\\
$b_{11}$ & $-4.013195 ~10^{-7}$\\
$b_{12}$ & $-2.1143606 ~10^{-6}$\\
$b_{13}$ & $-5.2677931~10^{-6} $\\
$b_{14}$ & $1.58535023 ~10^{-5} $\\
$b_{15}$ & $-1.305467659~10^{-4} $\\
$b_{16}$ & $1.1142 ~10^{-9} $\\
$b_{17}$ & $5.0965 ~10^{-9} $\\
$b_{18}$ & $1.96269~10^{-8}$\\
$b_{19}$ & $-1.0807 ~10^{-8}$\\
$b_{20}$ & $-2.899707~10^{-7} $\\
$b_{21}$ & $1.9280990 ~10^{-6} $ \\ 
\end{tabular}
\end{ruledtabular}
\end{table}

The dynamic viscosity of the aqueous sodium chloride solution as a function of temperature and composition at ambient pressure (1 bar) can be approximated by \cite{ofp1977}
\begin{eqnarray}
\mu(c,T)=&c_1+c_2e^{(a_1T)}+c_3e^{(a_2 m)} \nonumber \\
& +c_4e^{(a_3(0.01T+m))}+c_5e^{(a_4(0.01T-m))}
\label{eq:visco}
\end{eqnarray}
with $m=\frac{c/100}{(1-c/100)*M_{NaCl}}$ the molality of the solution. $M_{NaCl} = 58.44$~g\,mol$^{-1}$ is the molar mass of sodium chloride and $c$ (wt\%) is the concentration in salt in the solution.
The constants are defined as $c_1=0.1256735$, $c_2=1.265347$, $c_3=-1.105369$, $c_4=0.2044679$, $c_5=1.308779$, $a_1=-0.0429718$, $a_2=0.3710073$, $a_3=0.420889$, $a_4=-0.3259828$ \cite{ofp1977}.

\bibliography{biblio}

\end{document}